\documentclass[%
 reprint,
 amsmath,amssymb,
 aps,
floatfix,
]{revtex4-2}

\usepackage{url}
\usepackage[hyperindex,breaklinks]{hyperref}
\hypersetup{linkcolor=blue, citecolor=red, colorlinks=true}

\usepackage{braket}
\usepackage{physics}
\usepackage{mathtools}
\usepackage{amsfonts}

\usepackage{graphicx}
\usepackage{dcolumn}
\usepackage{bm}

\usepackage{float}

\usepackage{placeins} 

\usepackage{soul}

\begin{document}

\preprint{APS/123-QED}

\title{Quantum walks in two dimensions: controlling directional spreading with entangling coins and tunable disordered step operator}


\author{Caio B. Naves$^{1}$}
\thanks{caio.naves@usp.br}

\author{Marcelo A. Pires$^{2}$}
\thanks{piresma@cbpf.br}

\author{Diogo O. Soares-Pinto$^{1}$}
\thanks{dosp@ifsc.usp.br}

\author{S\'{\i}lvio M. \surname{Duarte~Queir\'{o}s}$^{3}$}
\thanks{sdqueiro@cbpf.br} 

\affiliation{
$^{1}$Instituto de F\'{\i}sica de S\~ao Carlos, Universidade de S\~ao Paulo, CP 369, 13560-970, S\~ao Carlos, SP, Brazil
\\
$^{2}$Departamento de Física, Universidade Federal do Cear\'a, 60451-970,  Fortaleza, Brazil
\\
$^{3}$Centro Brasileiro de Pesquisas F\'{\i}sicas, Rua Dr Xavier Sigaud 150, 22290-180 Rio de Janeiro - RJ, Brazil, and National Institute of Science and Technology for Complex Systems, Brazil
}

\date{\today}

\begin{abstract}

We study a 2-D disordered time-discrete quantum walk based on 1-D `generalized elephant quantum walk' where an entangling coin operator is assumed and which paves the way to a new set of properties. 
We show that considering a given disorder in one direction, it is possible to control the degree of spreading and entanglement in the other direction.
This observation helps assert that the random quantum walks of this ilk serve as a controllable decoherence channel with the degree of randomness being the tunable parameter and highlight the role of dimensionality in quantum systems regarding information and transport.
    

\end{abstract}

\maketitle


\section{\label{sec:intro}Introduction}

The establishment of the first (discrete-time) protocol for a quantum walker on a grid~\cite{aharonov1993quantum} allowed not only the theoretical description of experimental quantum phenomena like photon emission exchange in a superconducting cavity~\cite{meschede1985maser,brune1987maser} -- among so many other physical systems~\cite{kempe2003quantum} --, but played a crucial role in moving from the conceptualization of quantum computation~\cite{feynman1982simulating,shor1994algorithms} into its materialization~\cite{georgescu2014quantum,portugal2013quantum}.

Being understood as the quantum analogue to the random walk, the first quantum walk has given rise to a myriad of models the diffusion properties of which have been a leading feature in terms of surveying. Nonetheless, classical systems differ awfully from quantum systems; namely, the latter abides by the superposition principle, the statistical notion of pure and mixture states as well as the concept of entanglement.

Albeit those subjects have been intensively explored for uni-dimensional quantum walks little has been made regarding higher-dimension quantum walk systems, especially when disorder is introduced in the quantum walk.
Concerning standard quantum walks, the first systematic studies can be ascribed to~\cite{mackay2002quantum} where it was also shown that the introduction of disorder can lead to the emergence of a  classical distribution. It is worth heeding the appearance of classical-like spreading can be observed in 1D quantum walks as well~\cite{brun2003quantum,schreiber2011decoherence,montero2016classical,kovsik2006quantum,joye2011random,ahlbrecht2012asymptotic,pires2021negative}.
Moreover, by increasing the dimension of the quantum walk it is possible to increase the noise level and still obtain more coherence~\cite{oliveira2006decoherence}.

In all the aforementioned works it was employed a standard step operator; however, novel phenomena emerge when the discrete-time QW is employed with nonstandard translation operators~\cite{lavivcka2011quantum,zhao2015one,di2018elephant,pires2020quantum,pires2019multiple,das2019inhibition,sen2019unusual,sen2020scaling,mukhopadhyay2020persistent,zaman2022randomizing,pires2020quantum} or bespoken coins~\cite{venegas2005quantum,liu2009one,panahiyan2020one}.
For instance, in~\cite{di2018elephant} it was presented a  solvable quantum model, the elephant quantum walk (EQW), that is able to spread  hyperballistically.
Within this context, the introduction of a perpetual memory model~\cite{di2018elephant} -- which is poles apart to the markovianity of the canonical quantum walk -- has set forth the possibility of introducing a random quantum walk model yielding a broad diffusion behavior ranging from ballistic to hyperballistic~\cite{pires2019multiple}. Moreover, it was recently shown~\cite{naves2022} that such a class of quantum walks generates maximally entangled states for almost all initial coin states and coin operators whether the walker is at first localized or not.

That being said, in this paper we aim at probing the spreading and entanglement of a 2-D-quantum walker based on the generalized quantum walk protocol~\cite{pires2019multiple}. 
The latter property has been paid little attention to~\cite{annabestani2010asymptotic,yalccinkaya2015two,zeng2017discrete,angles2022quantum}.

The remainder of the manuscript goes as follows: In Sec.~\ref{sec:results} we present the results and in Sec.~\ref{sec:remarks} the discussion and the concluding remarks of our study.

\section{\label{sec:model}Model}
\subsection{\label{sec:standard_dqtw}The standard 2-D discrete time quantum walk}

Coined discrete time quantum walks correspond to the evolution of a quantum system, 
the \emph{quantum coin},
in a discrete position space, $\mathcal{H}_p$, through the association 
between its degree of freedom and a direction of motion. That evolution takes place by considering the action of a \emph{coin operator} updating the coin state --
like the classical random walk coin tossing -- and then the action of a \emph{shift operator} that changes the position state of the quantum walker accordingly. In the case
of the one-dimensional lattice 
$\mathcal{H}_p \equiv \mbox{span}(\{\ket{x} |\; x \in \mathbb{Z}\})$ and the quantum coin is a simple two-level system 
$\mathcal{H}_c \equiv \mbox{span}(\{\ket{\uparrow},\ket{\downarrow}\})$, with the unitary evolution given by
\begin{equation}
    U \equiv S(\mathbb{I}_p \otimes C_2)\mbox{ ,}
    \label{eq:unitary_op}
\end{equation}
where $S$ is the shift operator
\begin{equation}
    S \equiv \sum_{x \in \mathbb{Z}} \left( \ketbra{x + 1}{x}\otimes \ketbra{\uparrow}{\uparrow}
                                +   \ketbra{x - 1}{x}\otimes \ketbra{\downarrow}{\downarrow}
                                \right)\mbox{ ,}
    \label{eq:1Dshift_op}
\end{equation}
and $C_2$ the quantum coin toss operator
\begin{equation}
    C_2 (\theta,\beta,\gamma) \equiv   \begin{pmatrix}
                \cos\theta & \sin\theta e^{i\beta} \\
                \sin\theta e^{i\gamma} & -\cos\theta e^{i(\gamma + \beta)}
            \end{pmatrix}\mbox{ .}
    \label{eq:u2_coin}
\end{equation}
Instances of prevalent quantum coins operators are the \emph{Hadamard} with 
$(\theta = \pi/4,\beta = \gamma = 0)$,
\begin{equation}
    C_2(\pi/4, 0, 0) \equiv H = \frac{1}{\sqrt{2}}\begin{pmatrix}
                                                1 & 1 \\
                                                1 & -1
                                             \end{pmatrix},
    \label{eq:hadamard_op}
\end{equation}
and the \emph{Kempe} operators where $(\beta = \gamma = \pi/2)$ , 
\begin{equation}
    C_2(\theta, \pi/2, \pi/2) \equiv C_k(\theta) = 
    \begin{pmatrix}
        \cos\theta & i\sin\theta\\
        i\sin\theta & \cos\theta
    \end{pmatrix}.
    \label{eq:kempe_coin}
\end{equation}

\begin{figure}
    \centering
    \includegraphics[scale = 0.7]{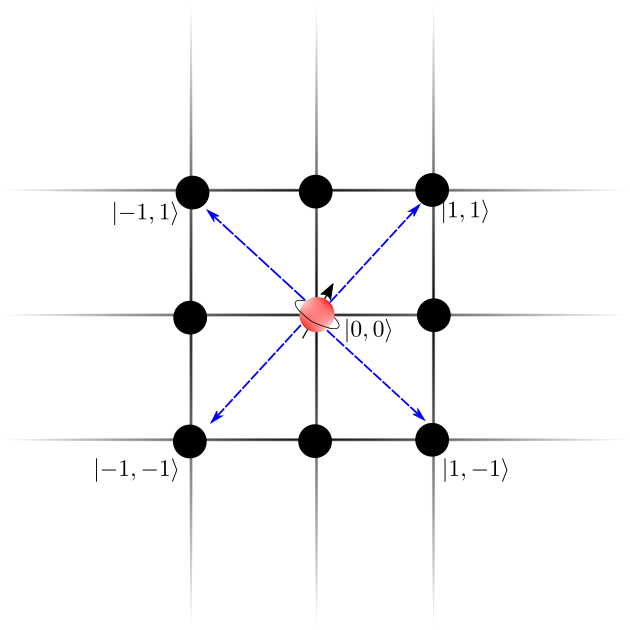}
    \caption{Representation of the 2-D coined discrete time quantum walk using a 
    composite quantum coin, made of two one-dimensional coins.}
    \label{fig:2d_dtqw}
\end{figure}

For higher dimensions, as the degree of freedom of the problem augments, it is possible to observe a wider range of ways in which the quantum system evolves. For example, in the two-dimensional lattice case, one can consider a qubit coin using it to move to one direction one step at a time, or to use different basis such as the eigenvectors of the Pauli matrices $\sigma_z$ and $\sigma_x$, to associate them to the directions of motion.
Here, we consider a four-dimensional coin made by a composition of two one-dimensional coins, that is, 
the coin state $\ket{\psi_c} \in \mathcal{H}_{c_1} \otimes \mathcal{H}_{c_2} = 
\mbox{span}(\{\ket{\uparrow,\uparrow},\ket{\uparrow,\downarrow},\ket{\downarrow,\uparrow},
\ket{\downarrow,\downarrow}\})$ such that the motion of the quantum walker occurs on the diagonals of the lattice (see Fig.~\ref{fig:2d_dtqw}).

As  we now have a composite coin, we can use two types of coin operators, the 
\emph{separable} ones, i.e.
    
    \begin{equation}
        C_S \equiv C^{(1)}_2 \otimes C^{(2)}_2\mbox{ ,}
        \label{eq:2d_sep_coin}
    \end{equation}
that acts individually in each subcoin, without creating correlations between 
them, or the \emph{non-separable} ones.

One example of a non-separable coin operator, that was used in this work, is the \emph{entangling coin operator}, where we use the controlled NOT gate operator (CNOT) to entangle the coins at each coin toss

    \begin{equation}
        C_{entang} \equiv \left(C^{(1)}_2 \otimes C^{(2)}_2 \right) 
        \begin{pmatrix}
            1 & 0 & 0 & 0 \\
            0 & 1 & 0 & 0 \\
            0 & 0 & 0 & 1 \\
            0 & 0 & 1 & 0
        \end{pmatrix}\mbox{ .}
        \label{eq:2d_entang_coin}
    \end{equation}


Making use of the same association between the coin states with the directions 
of motion in the one-dimensional case, the shift operator in the DTQW over a 2-D 
lattice is

    \begin{align}
        \displaystyle S = \sum_{x_1,x_2}\sum_{i,j} 
        &\ketbra{x_1 \pm 1,x_2 \pm 1}{x_1,x_2}\otimes
        \ketbra{i,j}{i,j}
        \mbox{ ,}
        \label{eq:2d_shift_op}
    \end{align}
where $i,j = \{\uparrow, \downarrow\}$ and the plus (minus) sign is selected 
when $i$ or $j$ is equal to the up (down) arrow.

\subsection{2-D generalized elephant quantum walk}

The generalized elephant quantum walk~\cite{pires2019multiple} (gEQW) is a unitary randomly 
modified version of the one-dimensional discrete time quantum walk with a non-standard 
shift operator, where the step sizes are drawn from a probability distribution. 
In this way, the shift operator is changed to

\begin{equation}
    S_t = \sum_{x \in \mathbb{Z}}\left(\ketbra{x + \Delta_t}{x} \otimes 
          \ketbra{\uparrow}{\uparrow} + \ketbra{x - \Delta_t}{x} \otimes
           \ketbra{\downarrow}{\downarrow}\right)\mbox{ .}
    \label{eq:1Dgeqw_shift}
\end{equation}
where $\Delta_t$ is the step size selected accordingly with the $q$-exponential
distribution\cite{tsallis2009introduction}

\begin{equation}
    \mbox{Pr}(\Delta_t) \equiv e_q(\Delta_t) = \tau_t [1 - (1-q)\Delta_t]^{1/(1-q)}
    \label{eq:qexp}\mbox{ ,}
\end{equation}
with support given by
\begin{equation}
    \mbox{supp}(e_q(x)) =   \begin{cases}
                                [0, \frac{1}{1-q})\mbox{, }q \le 1\\
                                [0 , \infty)\mbox{, }q > 1\mbox{ .}
                            \end{cases}
    \label{eq:supp_qexp}
\end{equation}
where $q \in [0,\infty)$, $\Delta_t \in [1,\dots,t]$ and $\tau_t$ is a 
time-dependent normalization factor. 

\begin{figure}[!ht]
        \centering
        \includegraphics[scale = 0.315]{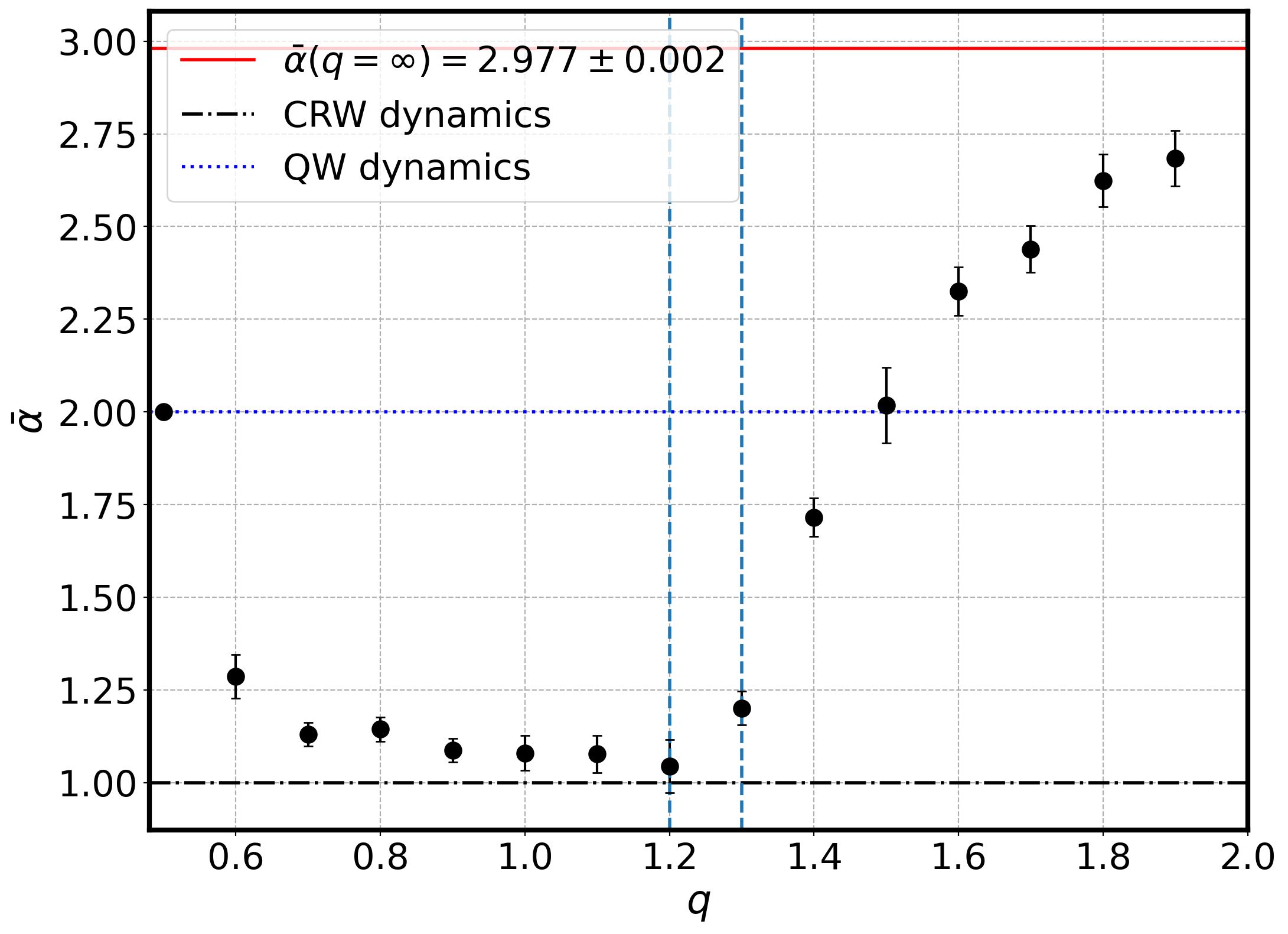}
        \caption{Average mean dynamical exponent of the one dimensional generalized elephant 
        quantum walk as a function of $q$ in the $q$-exponential distribution 
        Eq. (\ref{eq:qexp}). The vertical lines delimit the interval in which the average mean dynamical exponent starts to increase going to the hyper-ballistic regime (After Ref.~\cite{naves2022}).}
        \label{fig:malpha_x_q}
    \end{figure}

The $q$-exponential provide us with a way to change the quantum walk from the standard DTQW where 
$\Delta_t = 1$ and the dynamical behavior is characterized by
\begin{equation}
    \mbox{Var}_x(t) \approx t^{\alpha}\mbox{, }t \gg 1\mbox{ ,}
\end{equation}
with the dynamical exponent $\alpha = 2$, 
when we set $q = 1/2$, to the elephant quantum walk~\cite{di2018elephant} with 
$q \rightarrow \infty$ giving an average $\alpha = 3$, where the step sizes probabilities are drawn from the uniform distribution (for more details on the behavior of the dynamical exponent as a function 
of $q$ see Fig.~\ref{fig:malpha_x_q}).

The two-dimensional generalized elephant quantum walk model we propose is a modified version of the 2D DTQW described previously. Here, the shift operator Eq.~(\ref{eq:2d_shift_op}) is similar to the one-dimensional version but we must consider two step size distributions for each direction. For each time step the shift operator is
    \begin{align}
        S_t = \sum_{x_1,x_2}\sum_{i,j}
        &\ketbra{x_1 \pm  \Delta^{(1)}_t, x_2 \pm \Delta^{(2)}_t}{x_1,x_2}
        \otimes \ketbra{i,j}{i,j}\mbox{ ,}
        \label{eq:geqw_shiftop}
    \end{align}
where $\Delta_t^{(1)}, \Delta_t^{(2)}$ are selected independently and according  to the $q$-exponential distribution Eq. (\ref{eq:qexp}) and 
$i,j = \{\uparrow, \downarrow\}$ and the plus (minus) sign is selected when $i$ or $j$ is equal to the up (down) arrow. The coin operators used are the same as Eqs.~(\ref{eq:2d_sep_coin})~and~(\ref{eq:2d_entang_coin}).

\begin{figure}[!ht]
    \centering
    \includegraphics[scale = 0.5]{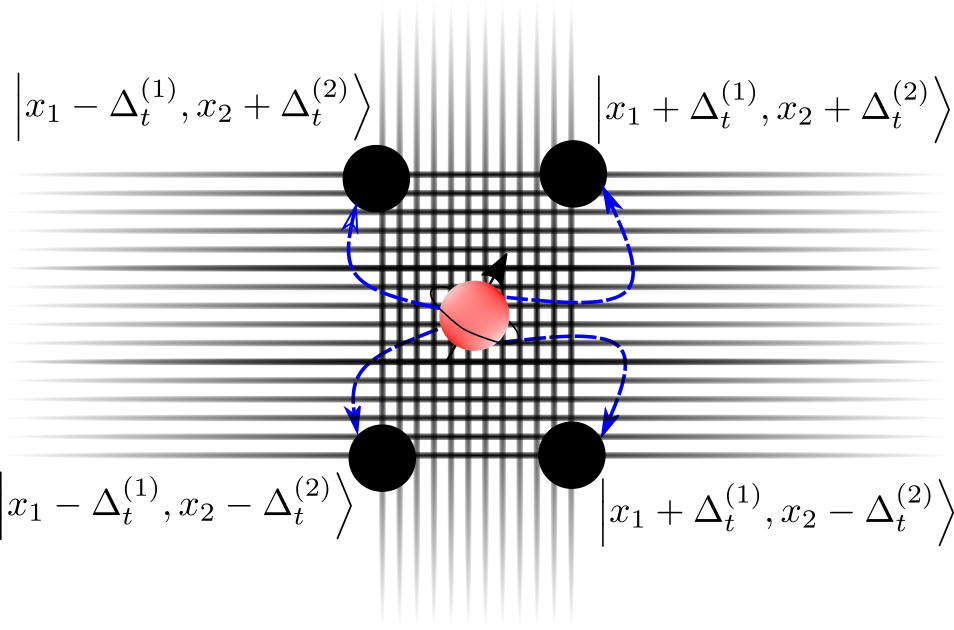}
    \caption{Representation of one step of the two-dimensional generalized 
    elephant quantum walk.}
\end{figure}

   \begin{figure}[!ht]
        \centering
        \includegraphics[scale = 0.5]{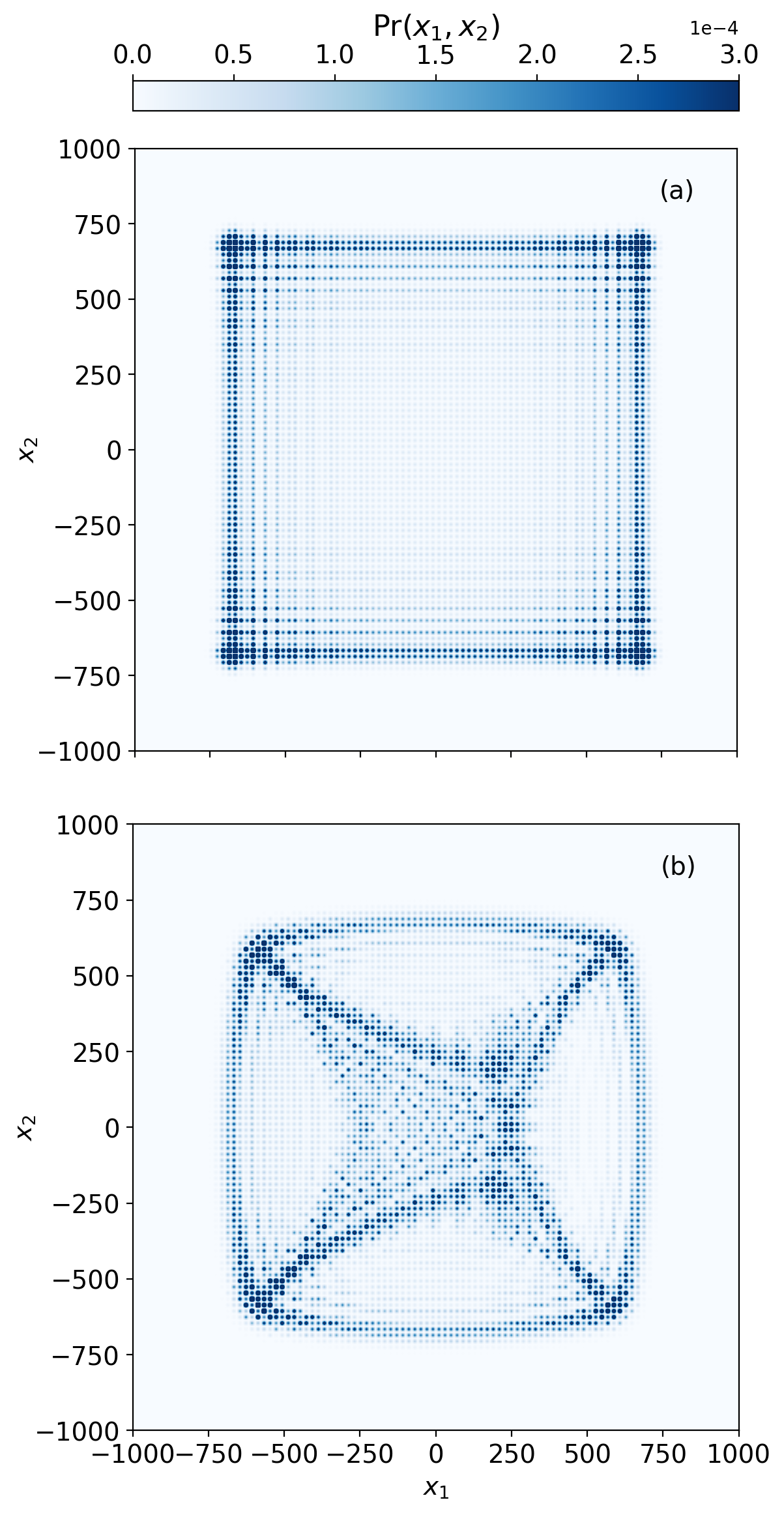}
        \caption{Position probability distribution for the two-dimensional 
        discrete time quantum walk with initial state 
        $\ket{\psi} = \ket{0,0}\otimes\frac{\ket{\uparrow} + 
        \ket{\downarrow}}{\sqrt{2}}\otimes
        \frac{\ket{\uparrow} + \ket{\downarrow}}{\sqrt{2}}$. In the top panel 
        (a), the coin operator used was Eq. (\ref{eq:2d_sep_coin}), with 
        $C^{(1)}_2$ and $C^{(2)}_2$ as the Kempe coin Eq. (\ref{eq:kempe_coin})
        with $\theta = \pi/4$. In the bottom panel (b), the entangling coin 
        operator was used, also with the same $C^{(1)}_2$ and $C^{(2)}_2$.}
        \label{fig:2d_stdqw_pd}
        \end{figure}

\section{Results}
\label{sec:results}

\subsection{Entangling 2-D generalized elephant quantum walk}
\label{subsec:e2dgeqw}

First, let us look at the probability distribution of the position in the standard DTQW on a two-dimensional lattice. In Fig.~\ref{fig:2d_stdqw_pd}(a), we have a discrete time quantum walk whose initial state is one localized in the origin and with both subcoins in the equal superposition state. The coin operator used was Eq.~(\ref{eq:2d_sep_coin}), with both coin operators, $C^{(1)}_2$ and $C^{(2)}_2$, as Eq.~(\ref{eq:kempe_coin}) with $\theta = \pi/4$. 
It is possible to learn that the region where it is more likely to find the quantum walker lies in the boundaries. That is an expected result bearing in mind the initial state and the coin operator used, as the evolution is separable for both directions. Notice that the 2-D unitary operator with separable coin operators can be written as 
$U = S_{x_1}C_{x_1}S_{x_2}C_{x_2}$, where 
\begin{eqnarray}
S_{x_1} &=& (\ketbra{x_1 + 1}{x_1} \otimes  \mathbb{I}_{x_2}\otimes \ketbra{\uparrow}{\uparrow} \otimes 
\mathbb{I}_{C_{x_2}} \\
&&+ \ketbra{x_1 - 1}{x_1}\otimes \mathbb{I}_{x_2} \otimes \ketbra{\downarrow}{\downarrow} \otimes \mathbb{I}_{C_{x_2}}),
\end{eqnarray}
\begin{eqnarray}
S_{x_2} &=& 
(\mathbb{I}_{x_1} \otimes \ketbra{x_2 + 1}{x_2} \otimes \mathbb{I}_{C_{x_1}} \otimes \ketbra{\uparrow}{\uparrow} \\
&&+ \mathbb{I}_{x_1} \otimes 
\ketbra{x_2 - 1}{x_2} \otimes \mathbb{I}_{C_{x_1}} \otimes 
\ketbra{\downarrow}{\downarrow}),
\end{eqnarray}
\begin{equation}
C_{x_1} = \mathbb{I}_{x_1,x_2} \otimes 
C_2^{(1)} \otimes \mathbb{I}_{C_{x_2}},
\end{equation}
and
\begin{equation}
C_{x_2} = \mathbb{I}_{x_1,x_2} 
\otimes \mathbb{I}_{C_{x_1}} \otimes C_2^{(2)}.
\end{equation}
Consequently, the quantum walk can be described as independent movements in the $\hat{x}_1$ and $\hat{x}_2$ directions and the two-dimensional probability distribution will be simply the product of two one-dimensional position probability distributions obtained from walks with the same individual initial states of both directions in the 2D DTQW.
(For more details, see Appendix~\ref{ap:sep})

    \begin{figure}[!ht]
        \centering
        \includegraphics[scale = 0.315]{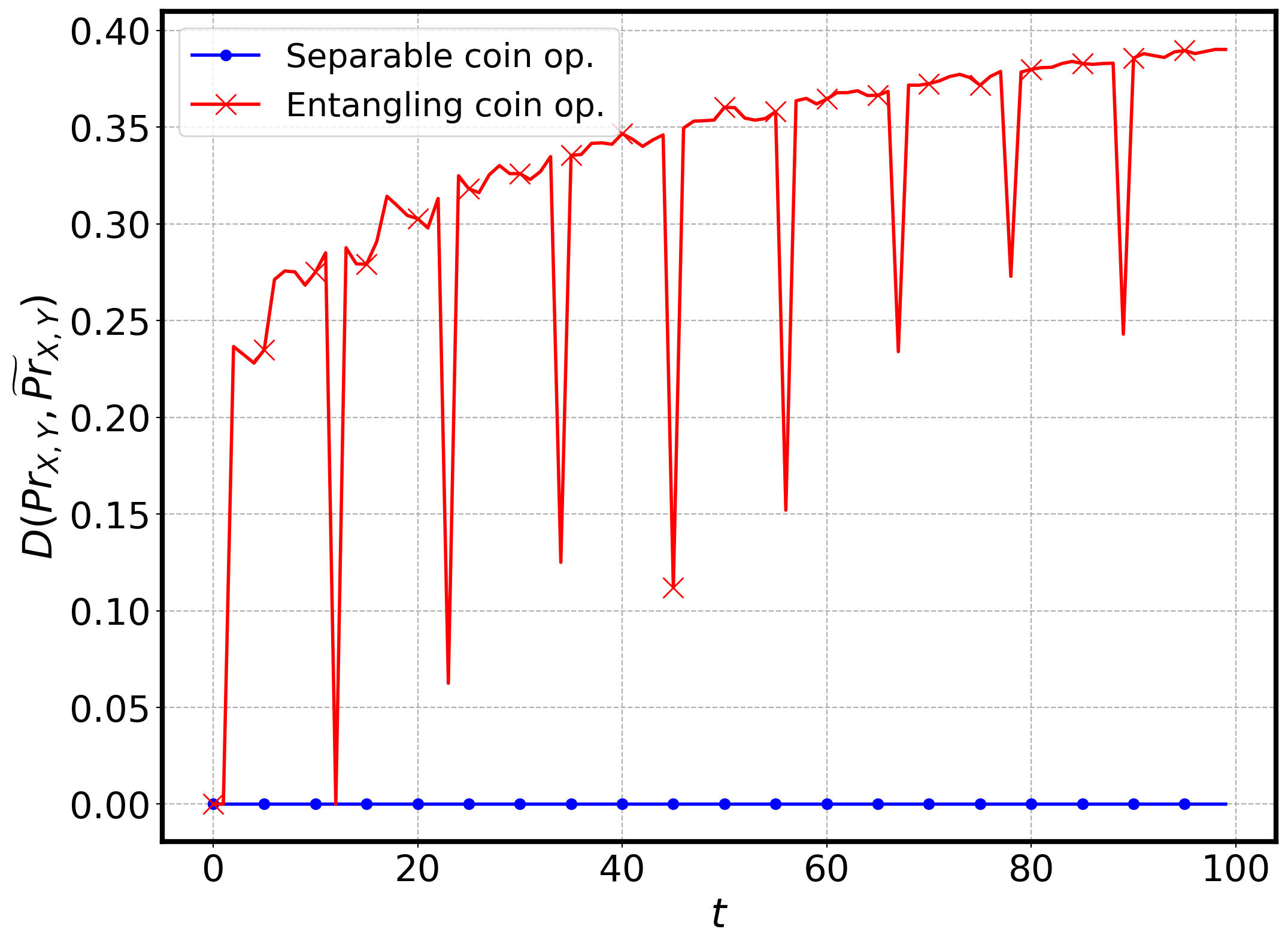}
        \caption{(Color online) Time series for the classical trace distance 
        between the joint distribution and the separable generated by its 
        marginals $\widetilde{\mbox{Pr}}_{X_1,X_2}$ = Pr$_{X_1}(x_1)$Pr$_{X_2}(x_2)$
        of the standard DTQW using the separable coin operator (blue circle) and 
        the entangling coin operator (red cross).
        The separable part of both coin operators walks was chosen as 
        $C_k(\pi/4) \otimes C_k(\pi/4)$, using a localized position with equal 
        superposition of the coin basis states as a walker initial state.}
        \label{fig:dtqw_td_entang}
    \end{figure}

    \begin{figure}[!ht]
        \centering
        \includegraphics[scale=0.315]{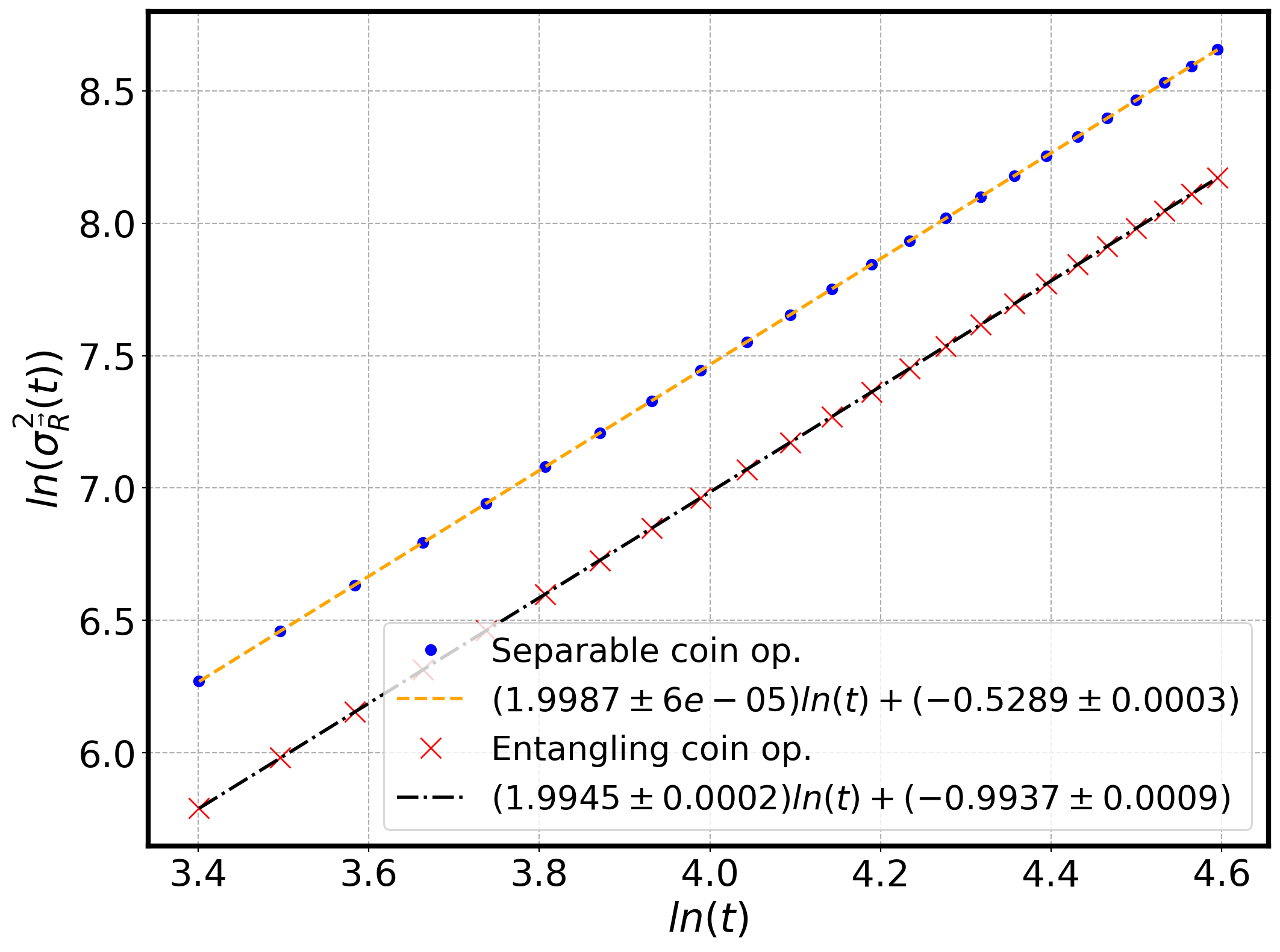}
        \caption{(Color online) Log-log graph of the position vector 
        variance for the standard DTQW on a 2D lattice, using the separable 
        coin operator Eq. (\ref{eq:2d_sep_coin})(circle blue color) and the 
        entangling coin operator 
        Eq. (\ref{eq:2d_entang_coin})(cross red color) and the same 
        parameters as in Fig.~\ref{fig:2d_stdqw_pd} considering only the asymptotic
        part of the evolution.}
        \label{fig:2dstd_var}
    \end{figure}

If we include the CNOT gate in the coin operator we get a very different 
probability distribution Fig.~\ref{fig:2d_stdqw_pd}(b). First, we see 
this distribution is not spatially symmetric, indicating that it is
not separable as in the previous case. In order to quantify the non-separability 
of the distribution, we employ the classical trace distance measure between the joint distribution Pr$_{X_1,X_2}(x_1,x_2)$ and a separable distribution generated by its marginals, $\widetilde{\mbox{Pr}}_{X_1,X_2} \equiv \mbox{Pr}_{X_1}(x_1)\mbox{Pr}_{X_2}(x_2)$,

    \begin{equation}
        D(Pr_{X_1,X_2},\widetilde{Pr}_{X_1,X_2}) = \frac{1}{2}
        \sum_{x_1,x_2}|\mbox{Pr}(x_1,x_2) - \widetilde{\mbox{Pr}}(x_1,x_2)|\mbox{ ,}
        \label{eq:cl_trace_dist}
    \end{equation}
so that if the distribution is separable the trace distance will return zero.
Figure~\ref{fig:dtqw_td_entang} shows the trace distance between Pr and $\widetilde{\mbox{Pr}}$ in the separable (blue circle) and the entangling DTQW (red cross). For the entangling case, it shows that the trace distance grows with time indicating that the distribution is not separable, something that does 
not happen when we use the separable coin operator.
Physically, this occurs because the introduction of the CNOT gate in the coin operator creates correlations between the subcoins at every time step that are then transferred to the position-coin system by the shift operator; that yields an inseparable position distribution for the walker in both directions. 

Analyzing the spreading behavior of the entangling walk as plotted in Fig.~\ref{fig:2dstd_var} through
the position vector variance time evolution
    \begin{equation}
        \sigma^2_{\vec{R}} = \langle \vec{R}^2\rangle - \langle \vec{R}\rangle^2
        = \sigma_{x_1}^2 + \sigma_{x_2}^2
        \mbox{ ,}
    \end{equation}
we see that the ballistic characteristic of the wavepacket spreading does not change.

    \begin{figure}[!ht]
        \centering
        \includegraphics[scale = 0.5]{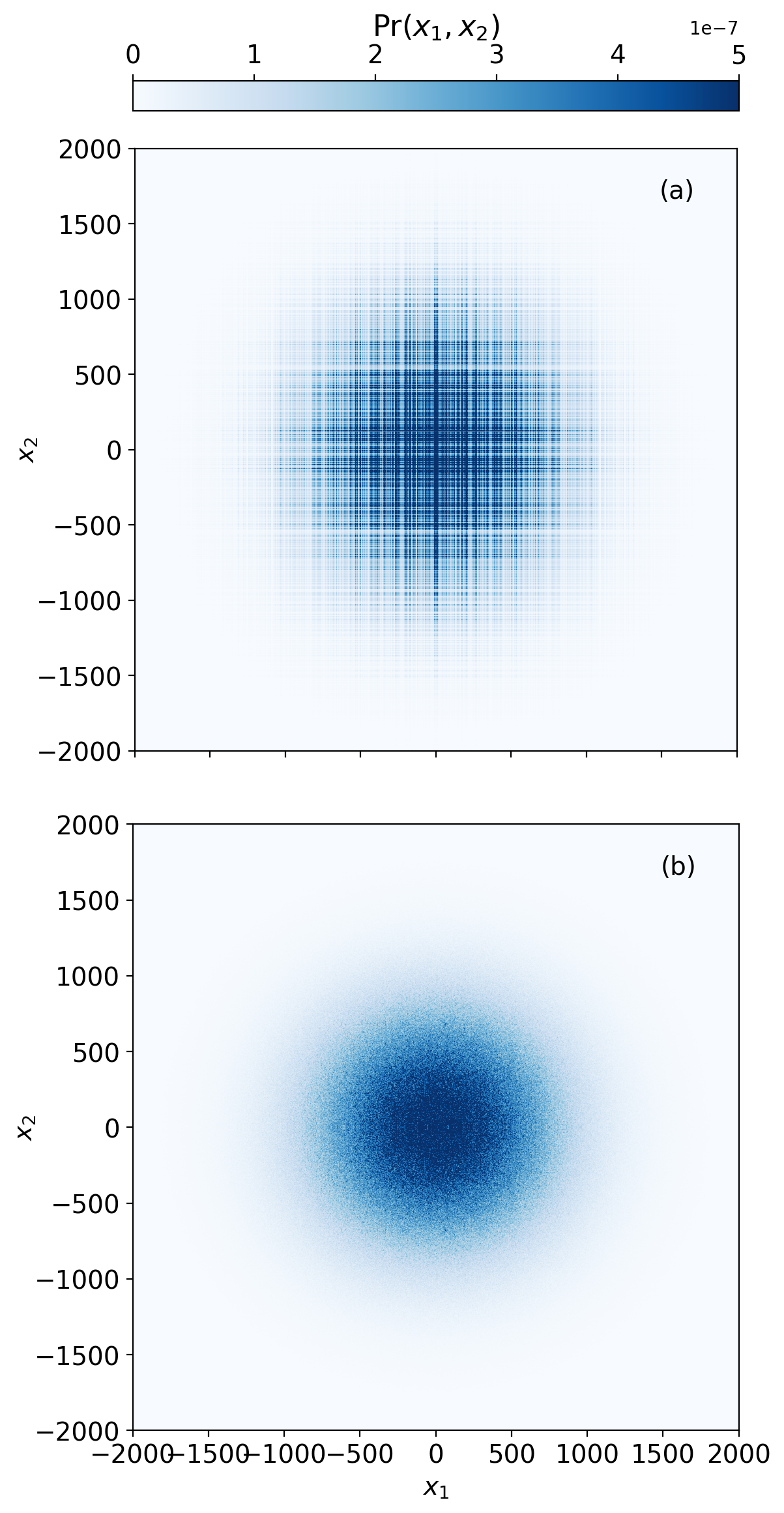}
        \caption{Position probability distribution for the two-dimensional 
        generalized elephant quantum walk using the $q$-exponential distribution 
        with $q = \infty$. The initial state used was 
        $\ket{\psi} = \ket{0,0}\otimes\frac{\ket{\uparrow} + 
        \ket{\downarrow}}{\sqrt{2}}\otimes\frac{\ket{\uparrow} + 
        \ket{\downarrow}}{\sqrt{2}}$. In the top panel (a), the coin operator 
        used was Eq. (\ref{eq:2d_sep_coin}), with $C^{(1)}_2$ and $C^{(2)}_2$ as 
        the Eq. (\ref{eq:kempe_coin}) with $\theta = \pi/4$. In the bottom panel 
        (b), the entangling coin operator was used, also with the same 
        $C^{(1)}_2$ and $C^{(2)}_2$.}
        \label{fig:2dgeqw_pd}
    \end{figure}

\begin{figure}[!ht]
        \centering
        \includegraphics[scale = 0.315]{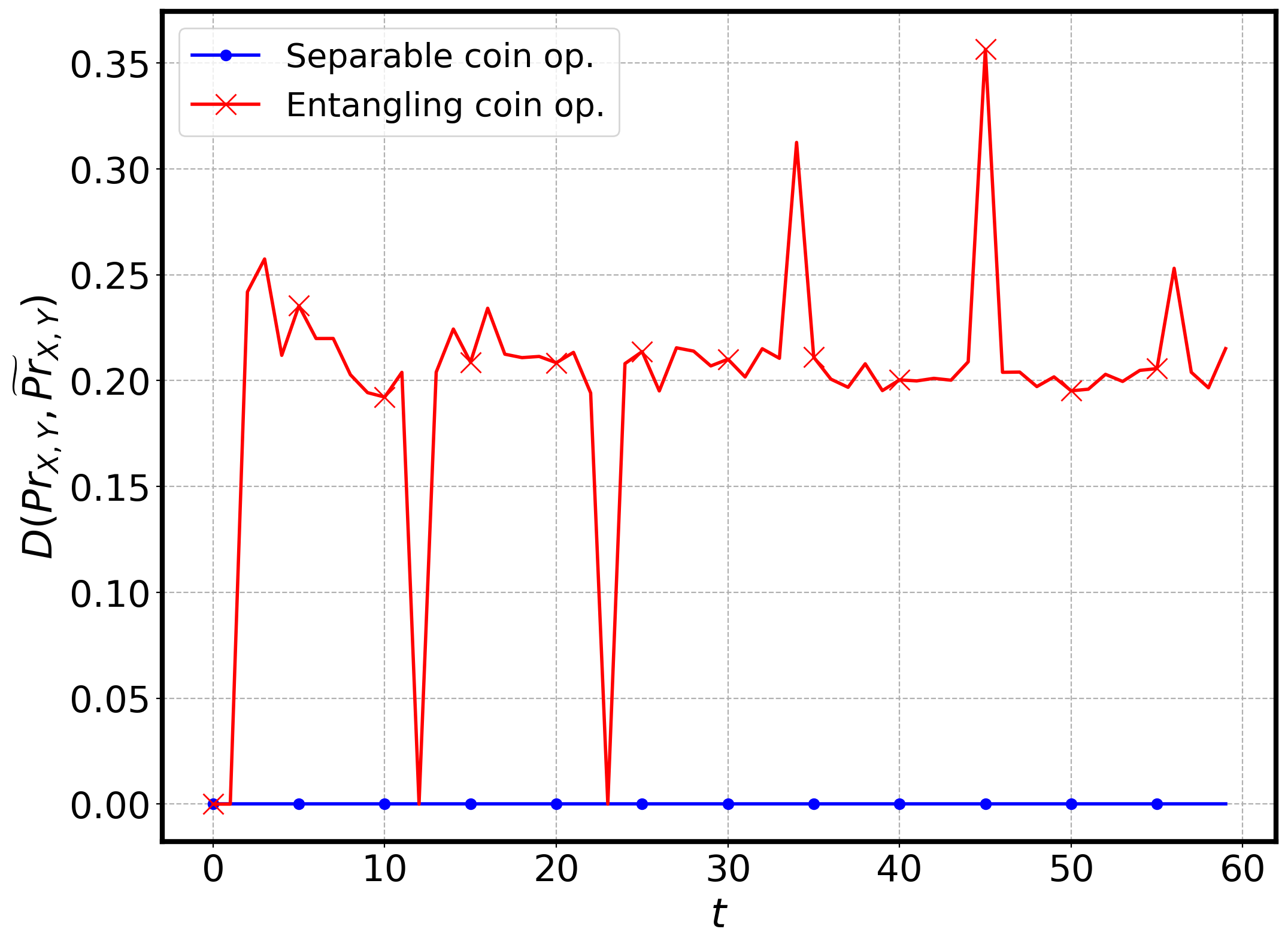}
        \caption{(Color online) Time series for the classical trace distance 
        between the joint distribution and the separable generated by its 
        marginals $\widetilde{\mbox{Pr}}_{X_1,X_2}$ = Pr$_{X_1}(x_1)$Pr$_{X_2}(x_2)$
        of the 2D EQW, $q_{x_1} = q_{x_2} = \infty$, using the separable coin 
        operator (blue circle) and 
        the entangling coin operator (red cross).
        The separable part of both coin operators walks was chosen as 
        $C_k(\pi/4) \otimes C_k(\pi/4)$, using a localized position with equal 
        superposition of the coin basis states as a walker initial state.}
        \label{fig:eqw_td_entang}
    \end{figure}

Next, we consider the two-dimensional gEQW. First, taking an evolution in which 
the step sizes are selected according to the uniform distribution, i.e. 
$q = \infty$ for both directions. Fig.~\ref{fig:2dgeqw_pd} shows us a 2D view 
of the position probability distribution for the evolution with the separable 
coin operator Eq. (\ref{eq:2d_sep_coin}) (a) and the entangling coin operator 
Eq. (\ref{eq:2d_entang_coin}) (b). The shape of the probability distribution 
gets a more definite bell shape when we use the entangling coin operator. 
Yet, we can not assert that the 2D distribution is completely Gaussian as it is also inseparable as depicted in Fig.~\ref{fig:eqw_td_entang}.

    \begin{figure}[!ht]
        \centering
        \includegraphics[scale = 0.315]{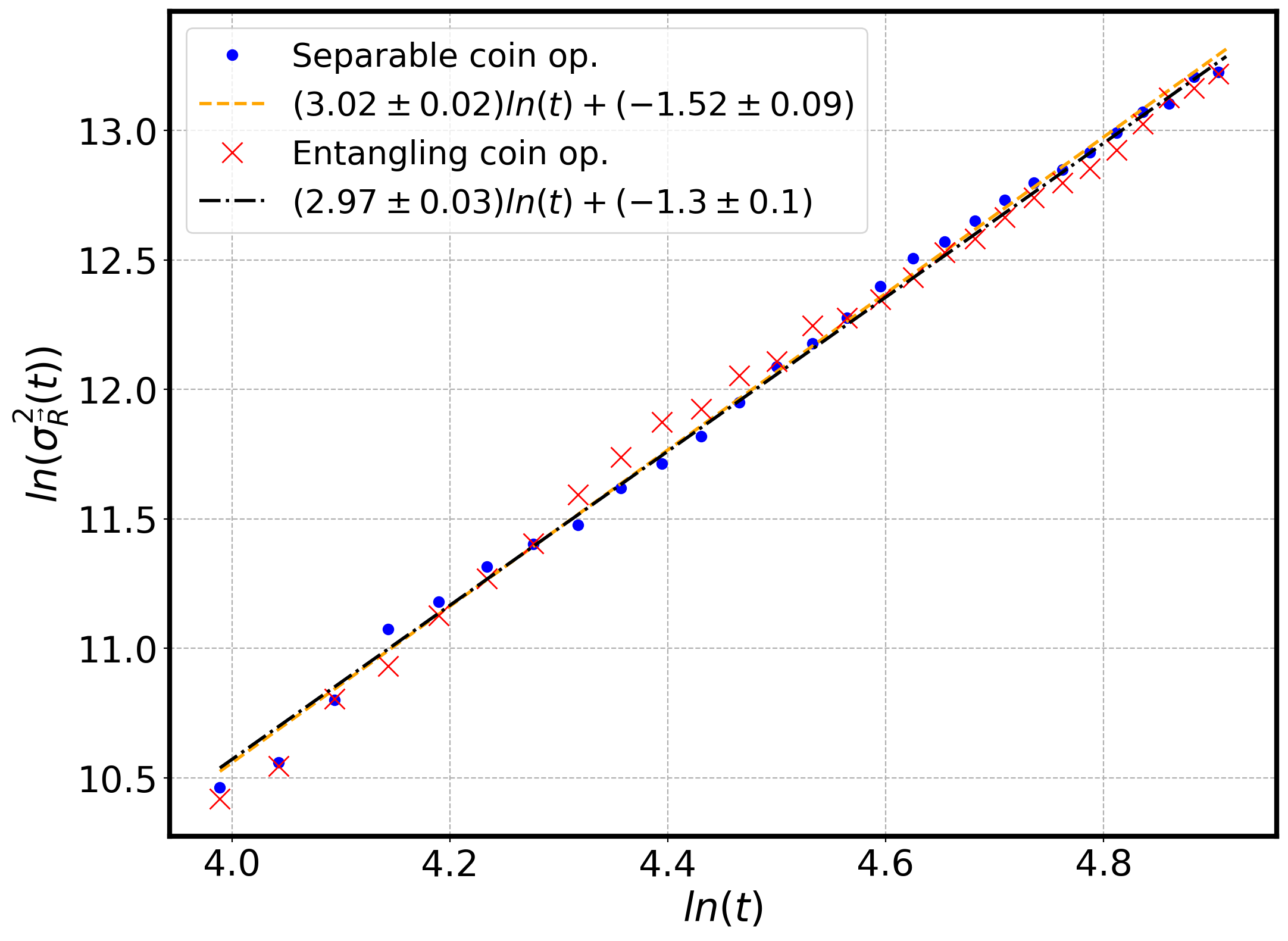}
        \caption{Log-log graph of the position vector variance as a function of 
        time for the 2D EQW in the asymptotic regime of Fig.~\ref{fig:2dgeqw_pd} 
        using the separable coin operator (blue circled and orange dashed lines) and the entangling coin 
        operator (red cross and black dash-dotted lines).}
        \label{fig:2deqw_var}
    \end{figure}

The position vector variance considering only the asymptotic regime is shown in the Fig.~\ref{fig:2deqw_var}. The results indicate that even when we use the entangling coin operator the dynamical exponent remains unchanged.

As a means of ulterior comparative analysis, we present the individual average dynamical exponents for each direction in Table~\ref{tab:mdif_eqw_entang}.

\begin{table}[!ht]
    \small
    \caption{Mean dynamical exponents in the 2D-gEQW and $q = \infty$ using entangling coin toss operation. These values were obtained through $10$ simulations in
        a $10^4 \cross 10^4$ lattice with a localized initial state and equal superposition of coin
        basis states were used. $C^{(1)}_2$ and $C^{(2)}_2$ were chosen as Eq. (\ref{eq:kempe_coin})
        with $\theta = \pi/4$.  }
        \label{tab:mdif_eqw_entang} 
        \begin{ruledtabular}
        \centering
        \begin{tabular}{ccc}
            & \mbox{full} & \mbox{asymptotic} \\
            \colrule
            $\bar{\alpha}_{x_1}$ & $2.82 \pm 0.02$ & $2.89 \pm 0.02$ \\
            $\bar{\alpha}_{x_2}$ & $2.77 \pm 0.02$ & $2.87 \pm 0.06$ \\
        \end{tabular}
        \end{ruledtabular}
\end{table}
\begin{table}[!ht]
        \caption{Mean dynamical exponents in the 2D-gEQW
        entangling coin toss operation with \textbf{$ \mathbf{q_{x_1} = 0.5}$} and \textbf{$\mathbf{q_{x_2} = \infty}$.}
        These values were obtained in $10$ simulations in a $10^4 \cross 10^4$ lattice, using 
        $\ket{\psi(0)} = \ket{0,0} \otimes \frac{\ket{\uparrow + \downarrow}}{\sqrt{2}} \otimes \frac{\ket{\uparrow + \downarrow}}{\sqrt{2}}$. $C^{(1)}_2$ and $C^{(2)}_2$ were chosen as Eq. (\ref{eq:kempe_coin}) with $\theta = \pi/4$.}
        \label{tab:mdif_05_inf_entang_table} 
        \begin{ruledtabular}
        \centering
        \begin{tabular}{ccc}
            & \mbox{full} & \mbox{asymptotic} \\
            \colrule
            $\bar{\alpha}_{x_1}$ & $1.040 \pm 0.004$ & $0.988 \pm 0.003$ \\
            $\bar{\alpha}_{x_2}$ & $2.78 \pm 0.02$ & $2.91 \pm 0.02$ \\
        \end{tabular}
    \end{ruledtabular}
\end{table}

Another type of evolution in the 2D-gEQW that we can carry out is by
considering different $q$ parameters in each direction. For example, we can take $q_{x_1} = 0.5$ and $q_{x_2} = \infty$. We expect that if we use the separable coin operator the results will be in accordance with the previous ones for the standard DTQW in one direction and the one-dimensional elephant quantum walk in the other, as there are no correlations between the walker's directions of movement. But what if we introduce correlations between the walker's coin subsystems? Recalling that the use of the CNOT gate together with the standard coin toss operation is a way to introduce correlations between the coins during the evolution, we proceed to the analysis of the 2D generalized elephant quantum walk with different $q$'s assigned for both directions. 

    \begin{figure}[!ht]
        \centering
        \includegraphics[scale = 0.5]{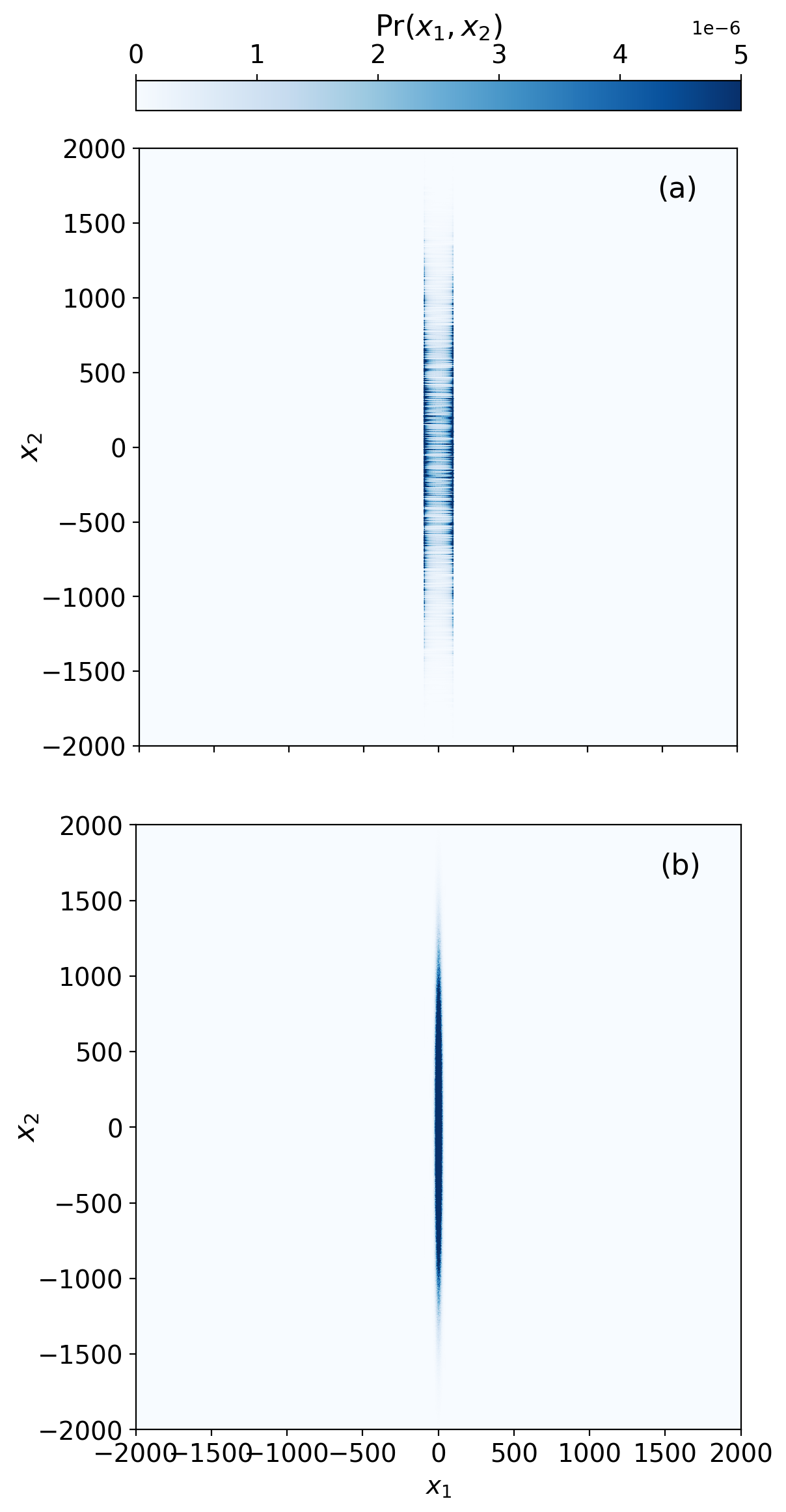}
        \caption{2D view of the position probability distribution for the 2D 
        gEQW. The parameters used were $q_{x_1} = 0.5$ for the first 
        direction and $q_{x_2} = \infty$ for the second. In the top panel 
        (a) the separable coin operator was used and in the bottom panel (b) 
        the entangling one. Both of them with 
        $C_2^{(1)} = C_2^{(2)} = C_k(\pi/4)$. The walker initial state was 
        localized in the lattice origin and with the equal superposition of the
        coin basis states.}
        \label{fig:2d_05_inf_pd_cmap}
    \end{figure}
    
    \begin{figure}[!ht]
        \centering
        \includegraphics[scale = 0.315]{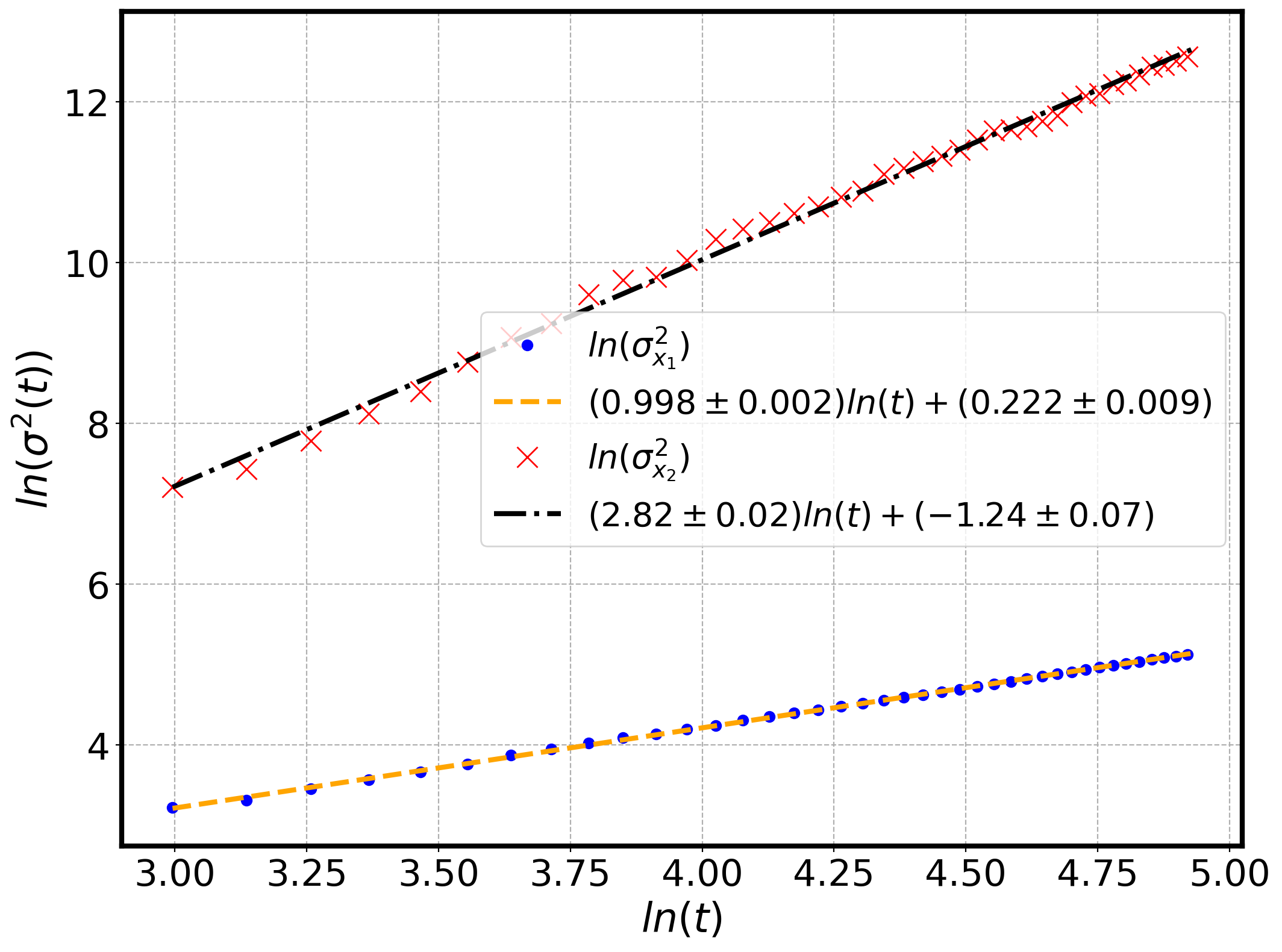}
        \caption{(Color online) Log-log graph as a function of time of the 
            $x_1$ position variance (blue circled) and $x_2$ variance 
            (red cross) with its linear
            fittings (orange dashed and black dash-dotted, respectively),
            in the 2D-gEQW  with $q_{x_1} = 0.5$ and $q_{x_2} = \infty$ of 
            Fig.~\ref{fig:2d_05_inf_pd_cmap}.}
        \label{fig:2d_05_inf_entang_var}
    \end{figure}

    

Figures~\ref{fig:2d_05_inf_pd_cmap}~and~\ref{fig:2d_05_inf_entang_var} depicts the results of a 2D-gEQW evolution with the entangling toss operation and $q_{x_1} = 0.5$ and $q_{x_2} = \infty$. Interestingly, we see that the $\hat{x}_1$ position distribution becomes much more localized with the wavepacket spreading exponent lowering to the classical random walk one. By comparing with the standard 2D entangling toss DTQW evolution, 
Fig.~\ref{fig:2dstd_var}, we can affirm that the introduction of the quantum walk with random steps sizes selected following a uniform distribution 
in one direction of the 2D entangling toss DTQW makes the walker's position 
probability distribution and wavepacket spreading in the other direction 
behaves similarly to a classical random walk.  We can also verify, by 
comparing these results with  Fig.~\ref{fig:2dgeqw_pd} and 
Tab.~\ref{tab:mdif_eqw_entang}, that this feature does not appear if we use 
$q = \infty$ in both directions, inasmuch as the wavepacket spreading exponents is still the same as by using the separable coin toss operation. 
Tab. \ref{tab:mdif_05_inf_entang_table} shows the average dynamical exponent for both directions in ten simulations.

    \begin{figure}[!ht]
        \centering
        \includegraphics[scale = 0.315]{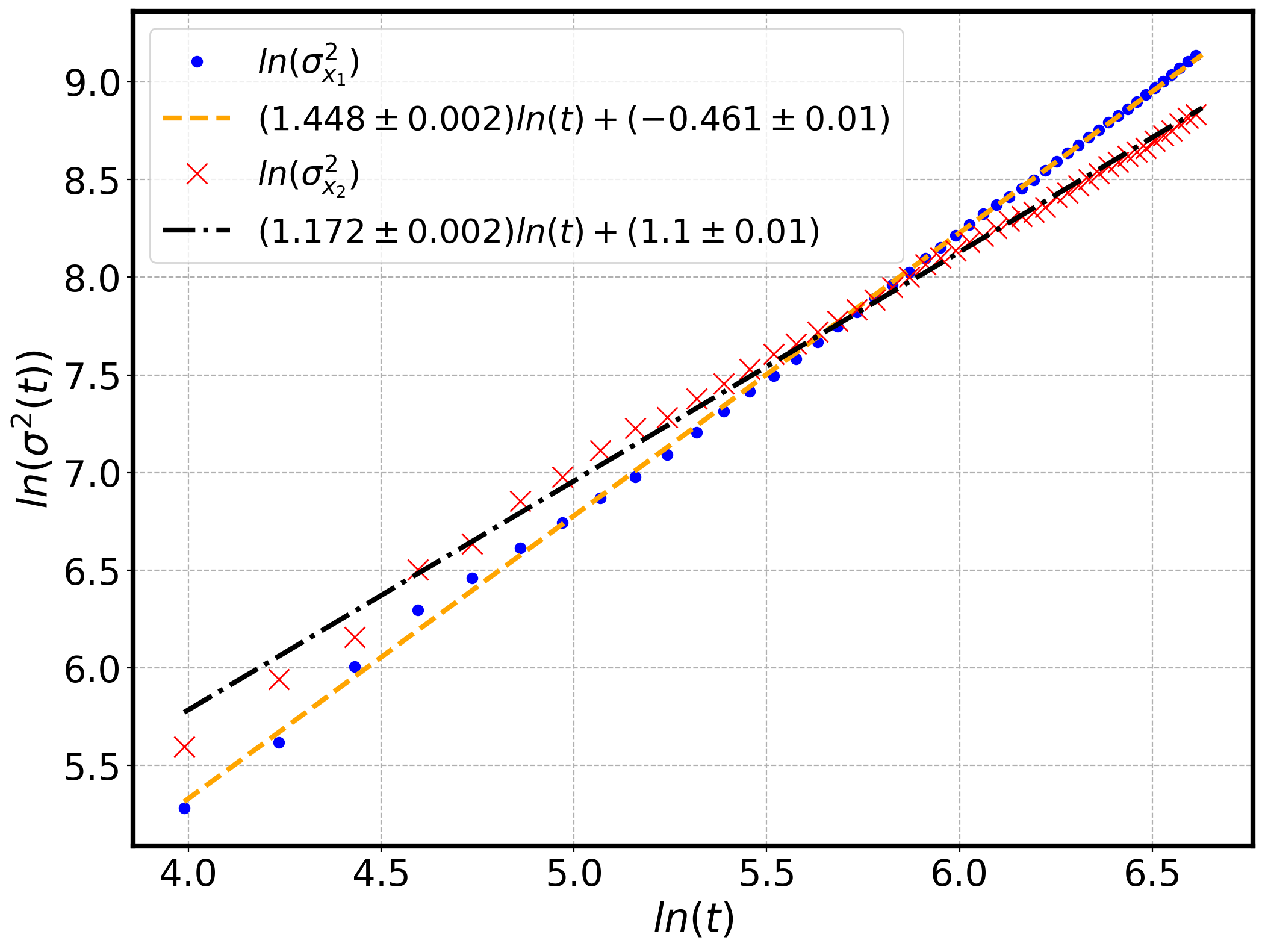}
        \caption{Log-log graph of the marginal variances and its linear 
        fittings, $\sigma_{x_1}^2$ (blue circle and orange dashed)
        and $\sigma_{x_2}^2$ (red cross and black dash-dotted) for the 
        $q_{x_1} = 0.5$ and $q_{x_2} = 1$ entangling 2D-gEQW in the asymptotic regime. 
        The initial state used was the same as in 
        Fig.~\ref{fig:2d_05_1_pd_cmap}. }
        \label{fig:2d_05_1_entang_var}
    \end{figure}
    
\begin{figure}[!ht]
        \centering
        \includegraphics[scale = 0.5]{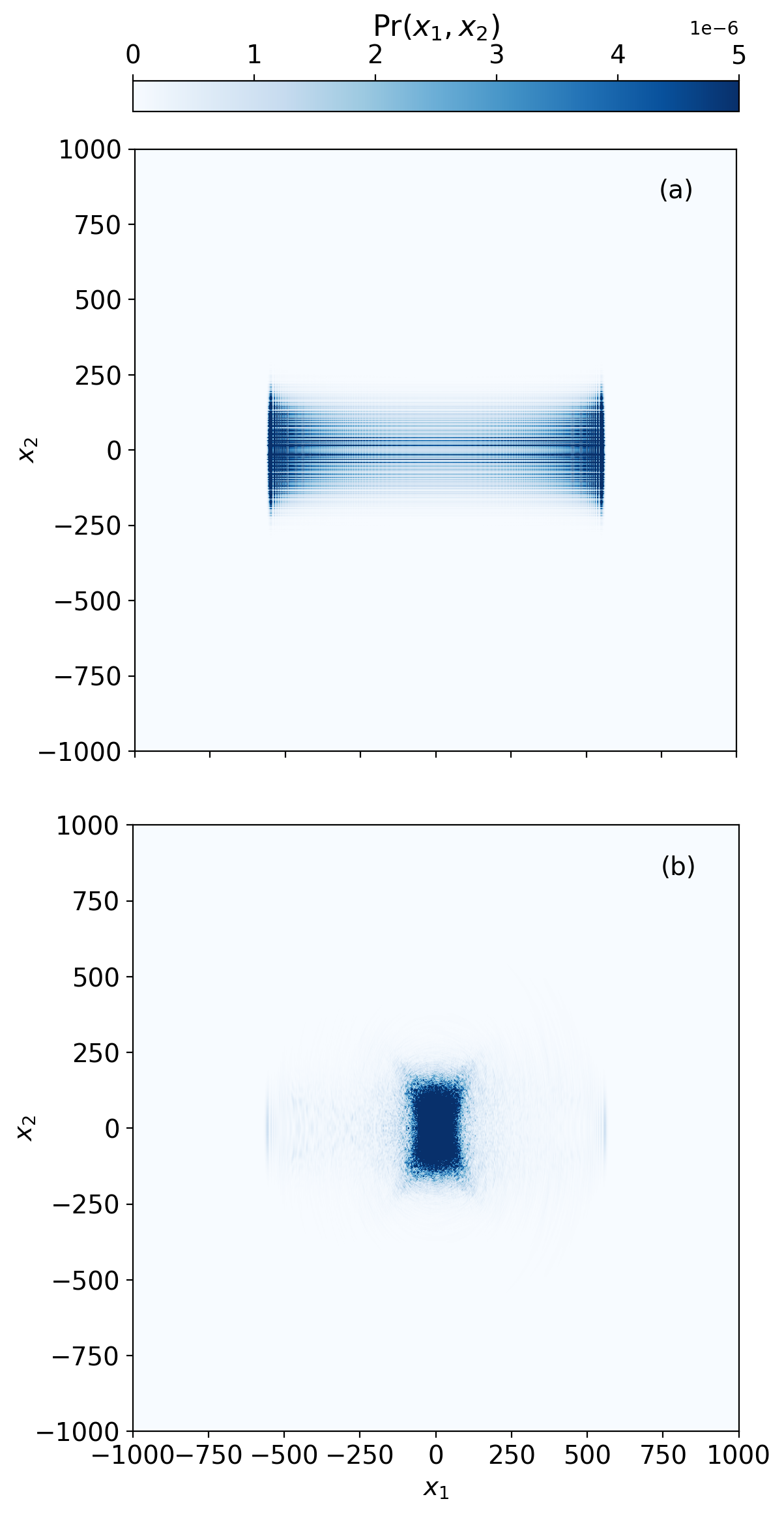}
        \caption{2D view of the position probability distribution for the 2D 
        gEQW. In the first direction the standard DTQW was used, i.e. with steps 
        with unit sizes and in the second the $q$-exponential distribution with 
        $q_{x_2} = 1$. The top panel (a) position distribution was obtained 
        using the separable coin operator Eq. (\ref{eq:2d_sep_coin}) with 
        $C_2^{(1)} = C_2^{(2)} = C_k(\pi/4)$ and the bottom panel (b) the 
        entangling one, with the same $C_2^{(1)}$ and $C_2^{(2)}$ as in the (a) 
        case.}
        \label{fig:2d_05_1_pd_cmap}
    \end{figure}

Changing the $q$-exponential distribution leads us to different results, 
nonetheless the wavepacket spreading suppression still occurs. 
Figs.~\ref{fig:2d_05_1_entang_var} and \ref{fig:2d_05_1_pd_cmap} shows the 
results for an evolution with $q_{x_1} = 0.5$ and $q_{x_2} = 1$. We perceive once again the probability distribution turns more localized in the $\hat{x}_1$ direction. Fig.~\ref{fig:2d_05_1_entang_var} tells us that the wavepacket spreading suppression occurred, but to a lower degree than in the previous case.
        
\begin{figure}[!ht]
        \centering
        \includegraphics[scale = 0.315]{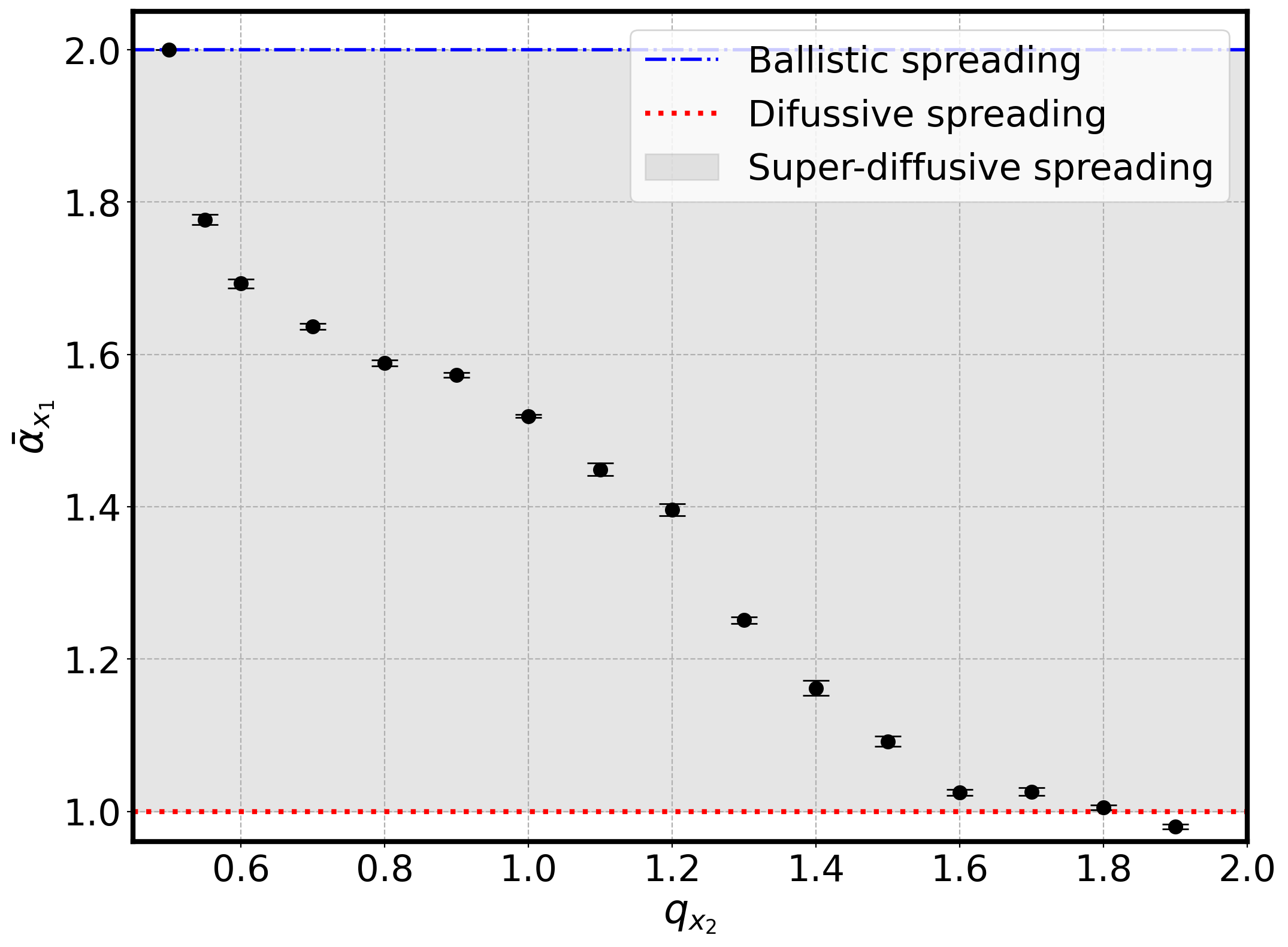}
        \caption{[MAIN RESULT] Average dynamical exponent of the first direction 
        variance with $q_{x_1} = 0.5$ as a function of $q_{x_2}$ of the gEQW in 
        the second direction. 
        The data error bars indicates the standard deviation of the points 
        obtained through $10$ simulations each. The 
        blue dashed-dotted line indicates the DTQW dynamical exponent and 
        the red dotted one the classical random walk dynamical exponent.
        In all simulations the localized and equal superposition of the basis 
        states was used, with $C_{entang} = C_k(\pi/4) \otimes C_k(\pi/4)
        \mbox{CNOT}$ as coin operator.}
        \label{fig:alphax1_x_qx2}
    \end{figure}

As a means to quantify the lowering degree of the quantum walker's first direction dynamical exponent as a function of the amount of randomness introduced in the second direction, we show in Fig.~\ref{fig:alphax1_x_qx2} the average dynamical exponent for the first direction $\bar{\alpha}_{x_1}$ as a function of $q_{x_2}$ in the entangling gEQW. We have determined that the average dynamical exponent has a non-monotonic decaying behavior as a function of $q_{x_2}$, increasing for some values when compared to the previous one. However, the overall behavior is to decrease as one increases the amount of randomness introduced on the the walk in the second direction, with a lower limit given by the classical random walk dynamical exponent $\alpha = 1$. \textcolor{black}{Differently from Ref.\cite{oliveira2006decoherence}, our direction-based protocol can be programmed to produce diffusive ($\alpha=1$) as well as superdiffusive ($\alpha>1$) spreading.}

\begin{figure}[!ht]
        \centering
        \includegraphics[scale = 0.325]{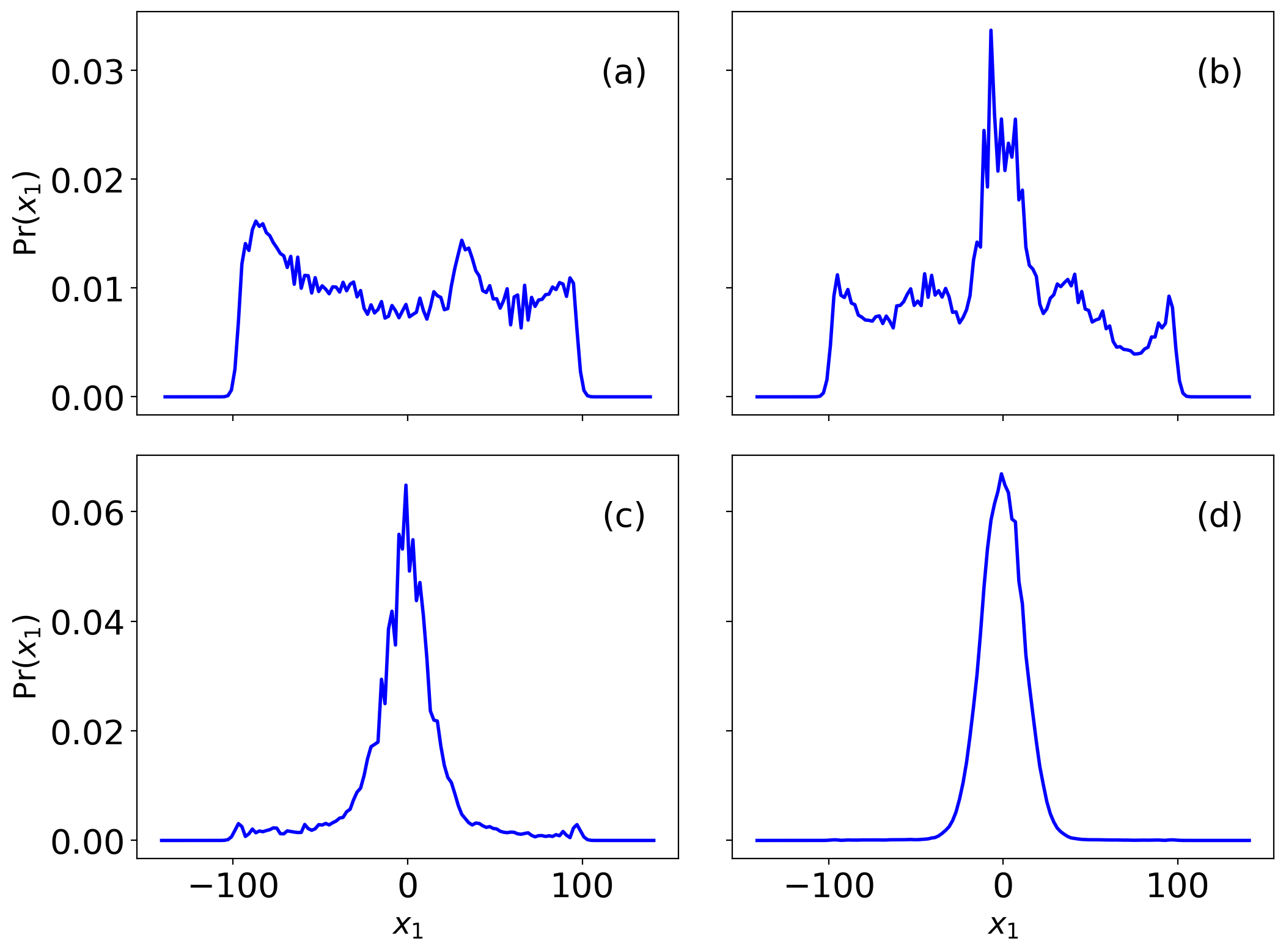}
        \caption{Marginal position probability distribution (without the zeros) 
        of the first direction for the gEQW using the entangling coin operator 
        Eq.(\ref{eq:2d_entang_coin}) with $q_{x_2} = 0.5$. In (a) we have the 
        standard DTQW, i.e. $q_{x_2} = q_{x_1} = 0.5$, (b) $q_{x_2} = 0.55$,
        (c) $q_{x_2} = 1$, (d) $q_{x_2} = \infty$, all in the same time step 
        $t = 140$. The walker initial state used in the one localized in the 
        origin with equal superposition of the basis states. The separable part 
        of the coin operator used was the Kempe coin with $\theta = \pi/4$.}
        \label{fig:pos_panel}
    \end{figure}

By looking at the marginal probability distribution for the walker in the first 
direction $\hat{x}_1$ we can gain further insight on what is happening. In 
Fig.~\ref{fig:pos_panel}(a) we have the standard DTQW using the entangling coin 
operator, showing that the distribution is not symmetric despite the use of an 
initially symmetric coin initial state. When increasing the amount of randomness 
in the second direction, for instance with $q_{x_2} = 0.55$ 
Fig.~\ref{fig:pos_panel}(b), the marginal distribution starts to be more 
localized around the original, yet with long tails resembling the 
one-dimensional DTQW distribution obtained when using a symmetric initial coin 
state. By further increasing $q_{x_2}$, the distribution becomes more Gaussian-like
with some remaining tail as in Fig.~\ref{fig:pos_panel}(c) for $q_{x_2} = 1$ 
and without it as in Fig.~\ref{fig:pos_panel}(d) with $q_{x_2} = \infty$ , where 
we see it as an almost completely Gaussian distribution, resembling the 
classical random walk distribution.

Despite the classical-like characteristic of the quantum walk discussed 
above for the entangling 2D-gEQW with $q_{x_2} = \infty$, that is the 
gaussianity of the distribution and the dynamical exponent equal to one, it is 
important to note that the final state of the walker in the $\hat{x}_1$ 
and $\hat{x}_2$ direction, coin plus position, is an entangled one and 
consequently does not have a classical analog, as we are going to see next.

   \begin{figure}[!ht]
        \centering
        \includegraphics[scale = 0.31]{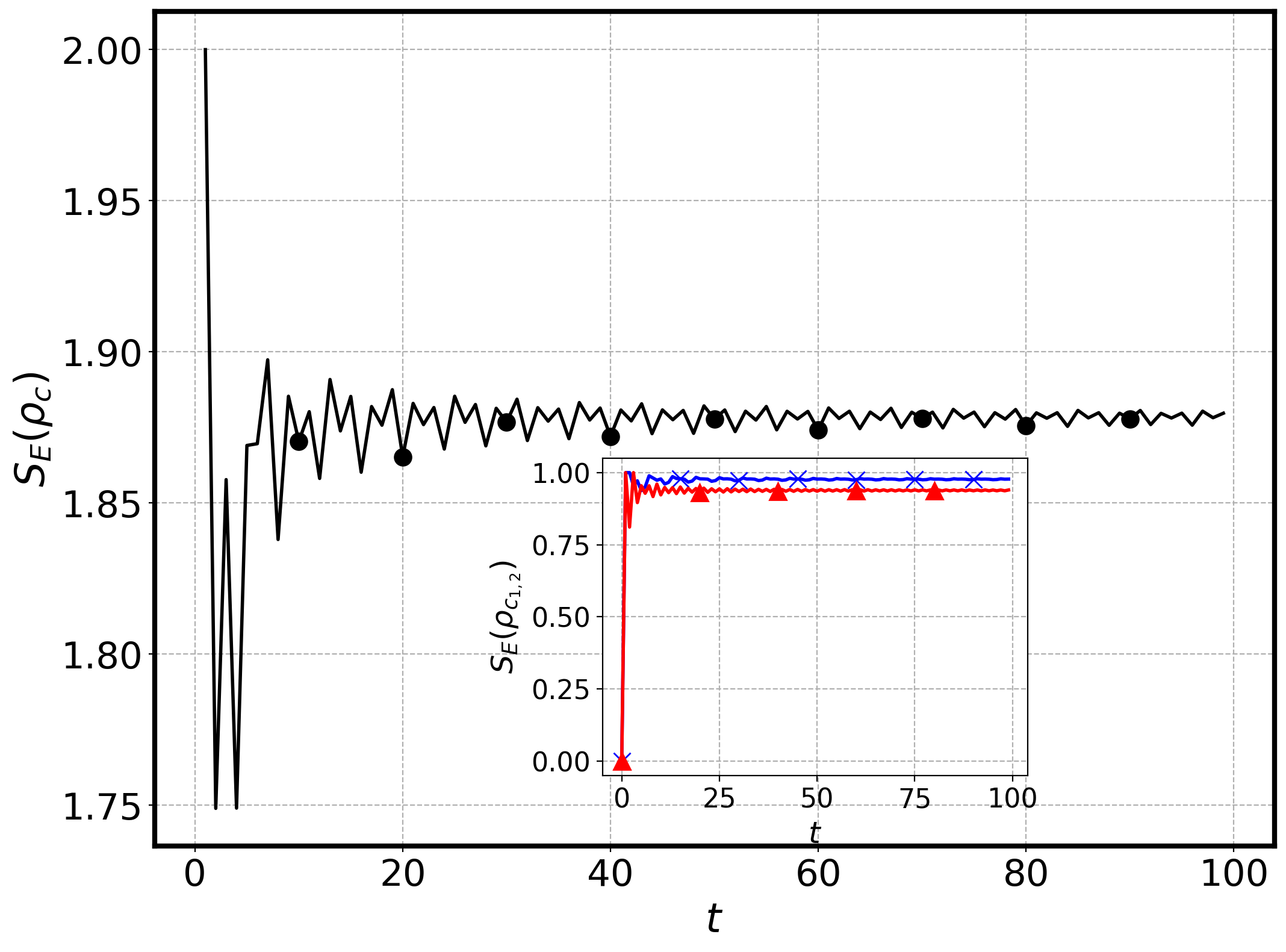}
        \caption{(Color online) Entanglement entropy of the total coin state in 
        the 2D entangling DTQW. The inset shows the entanglement entropy for the 
        first (blue cross) and second (red triangle) direction coin density 
        matrix. The separable part of the entangling coin operator used was with 
        equally balanced Kempe coin operators $C_k(\pi/4)$ and the initial state 
        used was the origin initially localized with equal superposition of the 
        coin basis states.}
        \label{fig:dtqw_entropy}
    \end{figure}
    
\subsection{Entanglement entropy}
\label{subsec:entang}

To assess the amount of entanglement of one part of the walker system has we are going to use the \textit{entanglement entropy} as quantifier~\cite{horodecki2009quantum}
    \begin{equation}
        S_E(\rho) = -\mbox{tr}(\rho \log_2 \rho)\mbox{ ,}
    \end{equation}
where $\rho$ is the density matrix of the system. The entanglement entropy has the property of having the lower limit equal to zero if the system 
is in a pure, and therefore separable, state and the maximum value equal to 
$\log d$, with $d$ being the system dimension, in the case of a completely 
entangled system.

Here we are going to consider three parts of the walker system, the total coin state $\rho_c$, the $\hat{x}_1$ and $\hat{x}_2$ coin states $\rho_{c_1}$ and $\rho_{c_2}$, respectively. Fig.~\ref{fig:dtqw_entropy} shows the entanglement entropy of the above mentioned coin systems in the entangling DTQW and show us that the two-dimensional coin has a non-zero entanglement with the two-dimensional position subsystem, going to the limit of $S_E(\rho_c) \approx 1.87$. The inset plot tells us that the $c_1$ subsystem is an almost maximal entangled state $S_E(\rho_{c_1}) \approx 0.97$, whereas for $c_2$ $S_E(\rho_{c_2}) \approx 0.94$. 
For comparison, in the one-dimensional Hadamard walk, the coin entanglement 
entropy reaches the asymptotic value of $S_E(\rho_c) \approx 0.87$\cite{carneiro2005entanglement,abal2006quantum}. This 
result supports our previous affirmation that the entanglement between the coin 
subsystems generated by the CNOT gate is transferred to the subcoin-position 
systems, since it gets a higher value than in the standard 1D DTQW.

\begin{figure}[!ht]
        \centering
        \includegraphics[scale = 0.31]{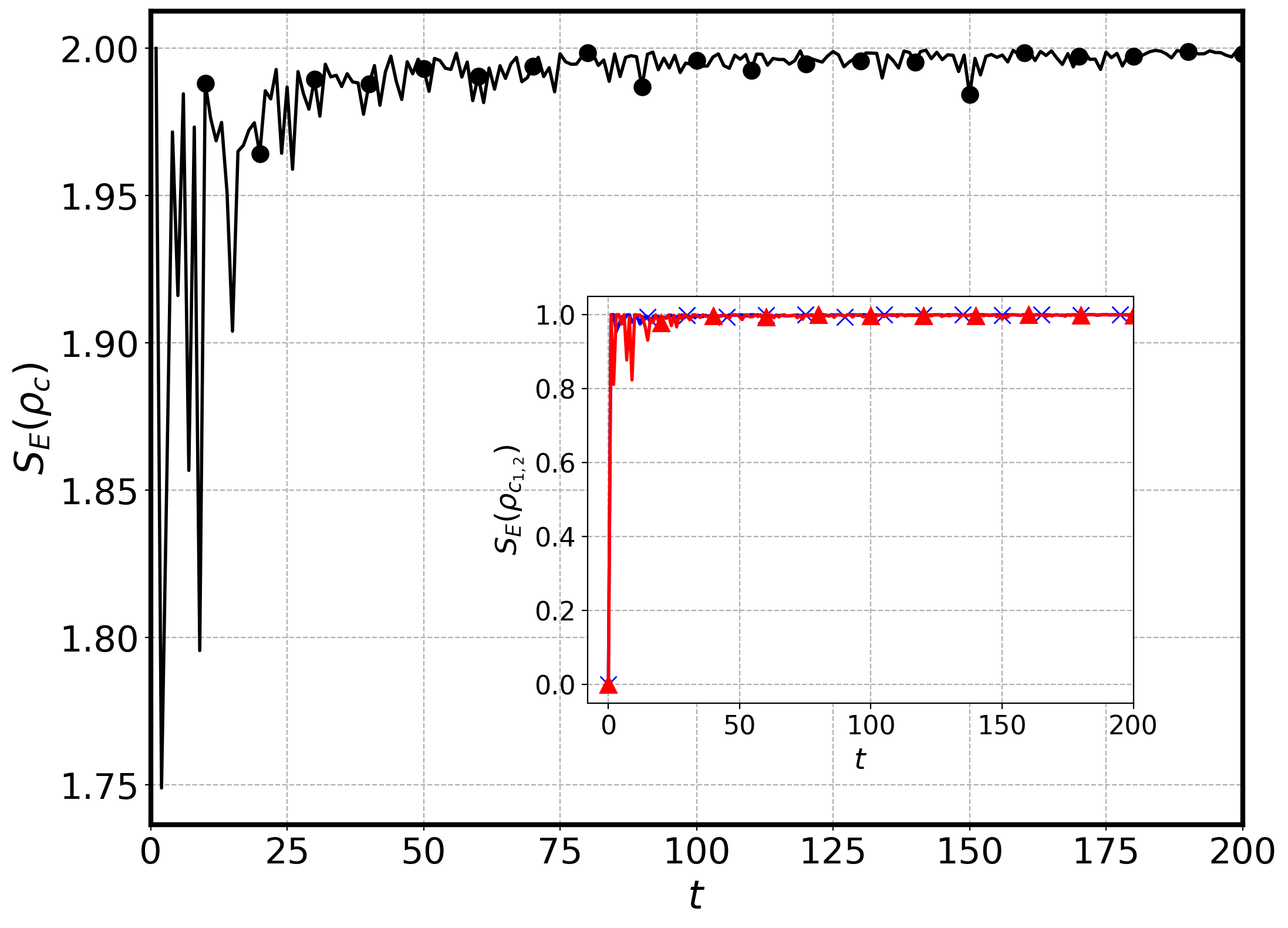}
        \caption{(Color online) Entanglement entropy of the total coin state in 
        the 2D-gEQW with $q_{x_1} = 0.5$ and $q_{x_2} = 1$. The inset shows 
        the entanglement entropy for the 
        first (blue cross) and second (red triangle) direction coin density 
        matrix. The separable part of the entangling coin operator used was with 
        equally balanced Kempe coin operators $C_k(\pi/4)$ and the initial state 
        used was the origin initially localized with equal superposition of the 
        coin basis states.}
        \label{fig:q_05_1_entropy}
    \end{figure}

Now taking into account the entangling generalized elephant quantum walk,  Fig.~\ref{fig:q_05_1_entropy}, with 
$q_{x_1} = 0.5$ and $q_{x_2} = 1$, we see that the 
total coin system becomes completely entangled with the position degrees of
freedom. Moreover, the inset shows that the coin's subsystems become also maximally entangled with their respective position degrees.

\begin{figure}[!ht]
        \centering
        \includegraphics[scale = 0.32]{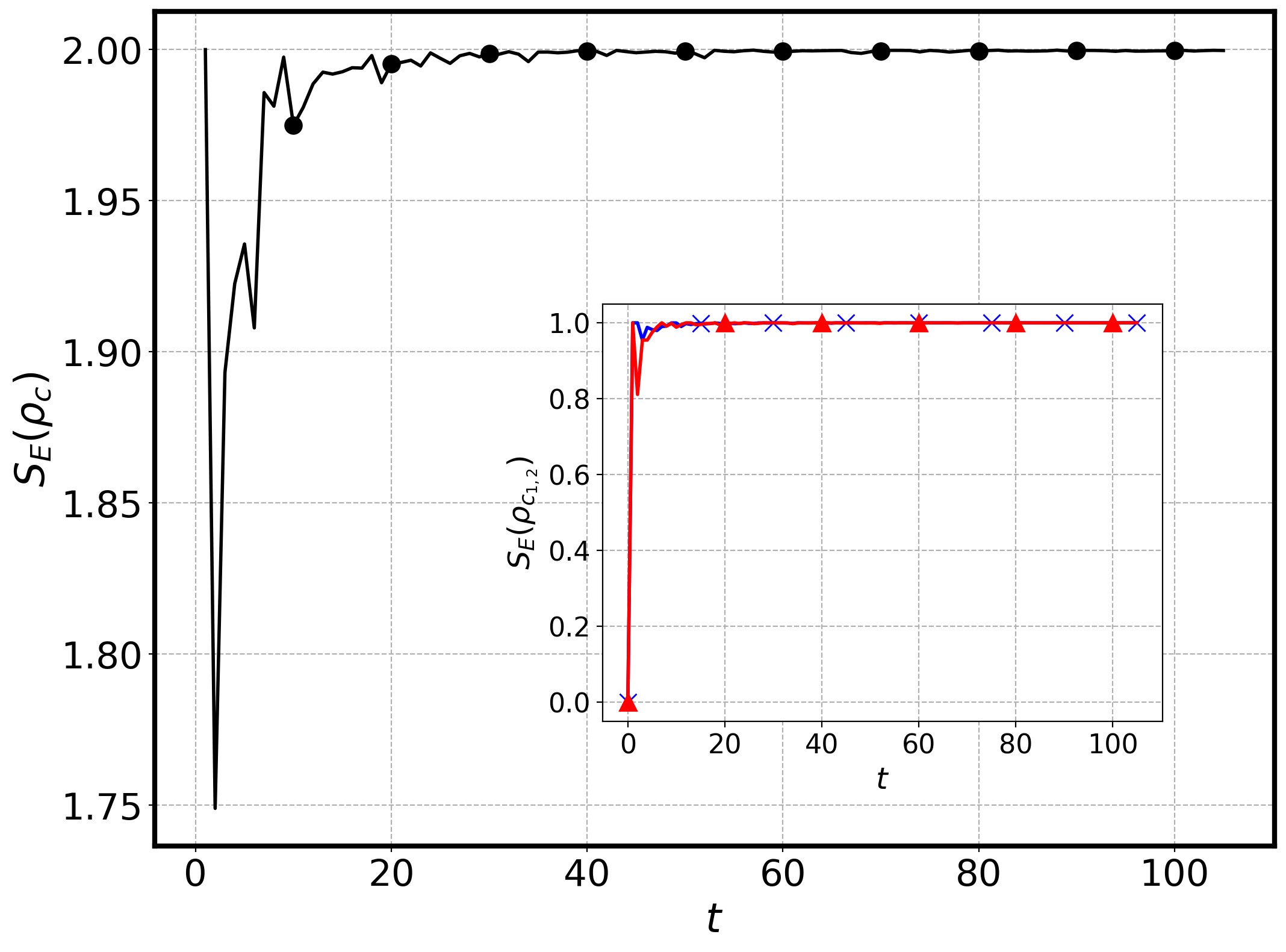}
        \caption{(Color online) Entanglement entropy of the total coin state in 
        the 2D-gEQW with $q_{x_1} = 0.5$ and $q_{x_2} = \infty$. The inset shows 
        the entanglement entropy for the 
        first (blue cross) and second (red triangle) direction coin density 
        matrix. The separable part of the entangling coin operator used was with 
        equally balanced Kempe coin operators $C_k(\pi/4)$ and the initial state 
        used was the origin initially localized with equal superposition of the 
        coin basis states.}
        \label{fig:q_05_inf_entropy}
    \end{figure}

When we use the uniform distribution of steps on the second direction, 
$q_{x_2} = \infty$, the same features appears Fig.~\ref{fig:q_05_inf_entropy}, 
i.e. a maximally entangled total 
coin state with its subsystems also maximally entangled with their respective 
position degrees. Therefore, as we mentioned previously, despite the quantum 
walk in the first direction have a Gaussian distribution with dynamical exponent 
equal to the classical random walk's, since it is maximally entangled with 
the position degree we can not say that it is a classical state.

One might question, how do we know if the entanglement of the coin subsystems is 
not between them? For answer that we use a measure of entanglement between 
bipartite systems that is called \textit{negativity}, $\mathcal{N}$\cite{asher1996,HORODECKI19961}. 
The negativity follows the PPT (positive partial transpose) criterion for 
separability which states that if the composite bipartite system parts $A$ and 
$B$ are entangled with each other, then the density matrix obtained through the 
partial transpose of $A$ or $B$ degrees of freedom has negative eigenvalues, 
violating the semi-definite positivity property. For $2 \cross 2$ systems the 
condition is also sufficient. It is defined as

    \begin{equation}
        \mathcal{N}(\rho_{AB}) = \frac{\|\rho_{AB}^{T_A}\|_1 - 1}{2}\mbox{ ,}
    \end{equation}
where $T_A$ is the partial transpose over $A$ and $\|.\|_1$ the trace norm. The 
negativity is an entanglement monotone, going from zero for separable states to 
half for completely entangled states. By calculating the negativity of $\rho_c$ 
for all the cases mentioned above, we see that the subcoins are not entangled 
with each other after the application of the unitary operators, as it returns 
zero for all times.

These results shows us that the 2-D generalized elephant quantum walk has a 
remarkable property, that we can control the degree of dispersion of the standard 
quantum walk in the first direction through the use of the entangling coin operator
while also enhancing the entanglement with its coin subsystem and maintaining the 
properties of the second direction. Taking into account the total coin system, the 
2-D entangling gEQW shows to be a walk capable of creating maximally correlated coin
states in a bidimensional setting.

\subsection{Physical interpretation}
\label{subsec:phys}

In order to understand the reason the walker spreading in the first direction is reduced in the 2-D gEQW with random step sizes in the second direction we have employed a normalized version of the $l_1$-norm coherence\cite{baumgratz2014,bera2015}.

\begin{equation}
    C_{l_1}(\rho) = \frac{1}{t}\sum_{j > i} |\rho_{i,j}|\mbox{ .}
\end{equation}
A normalization is made necessary since the position degree dimension is 
infinite, then as time passes the coherence would increase indefinitely given that more off diagonal elements would be non-zero.
The $l_1$-norm coherence provides us a way to see how the first direction position state coherence evolves through time and can potentially explain the previously mentioned properties of classical-random-walk-like behavior, since one expects that if the position density matrix is completely diagonal and one uses a equally balanced coin toss the evolution should lead to a Gaussian distribution.

\begin{figure}[!ht]
    \centering
    \includegraphics[scale = 0.32]{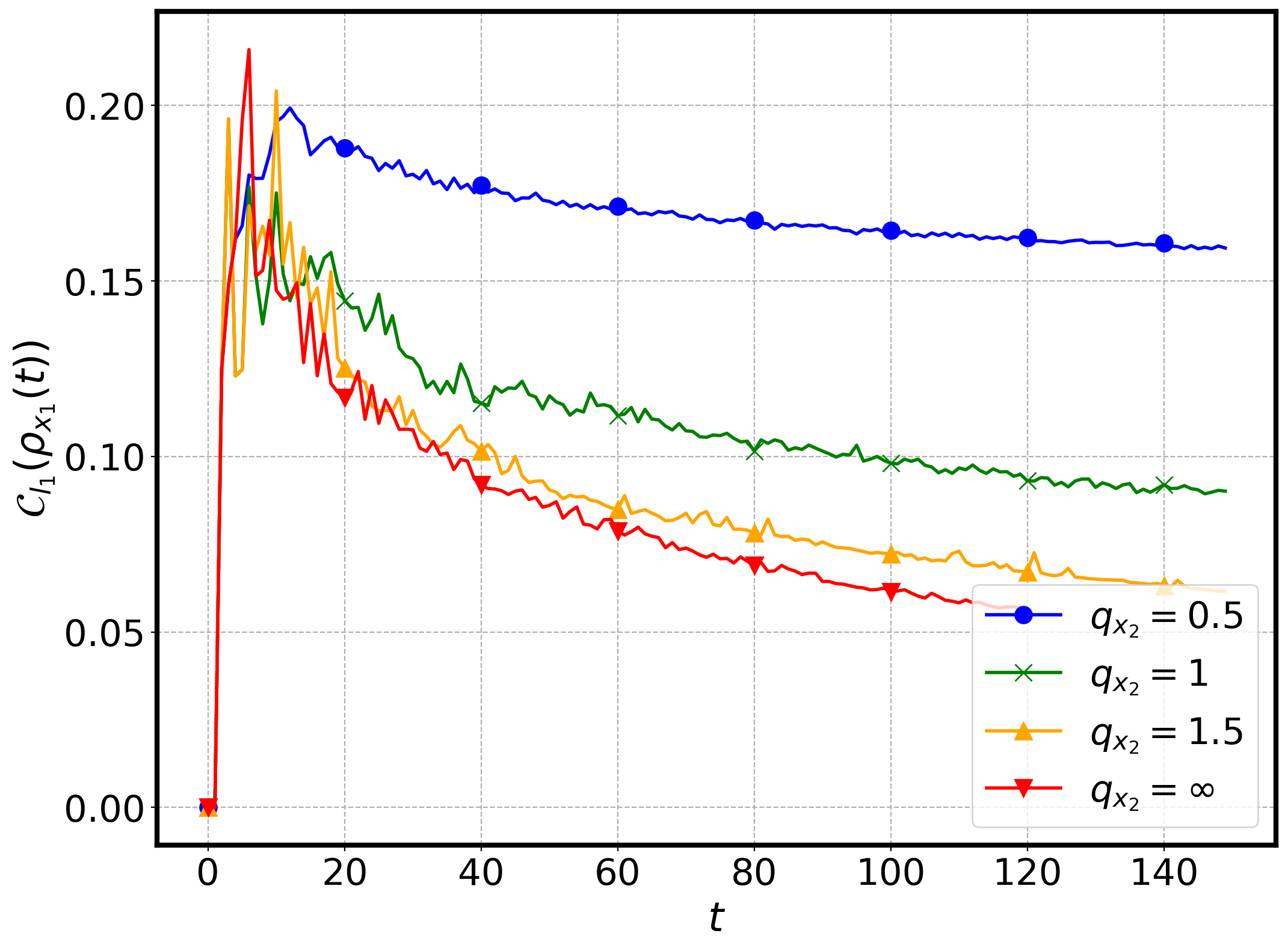}
    \caption{$l_1$-norm of coherence as a function of time in the entangling
    2-D gEQW with different values of $q_{x_2}$. In all simulations the Kempe coin operator was used with $\theta = \pi/4$ for both directions and the initially localized walker with an equal superposition of the coin basis states. Supposing that after the initial increase the coherence decays following a power law, $\mathcal{C}_{l_1} \propto t^{-\beta}$, a linear fitting of log-log of this graph give us the decay exponents as a function of  $q_{x_2}$, $\beta (q_{x_2} = 0.5) = (0.00744 \pm 0.00005)$, $\beta(q_{x_2} = 1) = (0.219 \pm 0.002)$, $\beta(q_{x_2} = 1.5) = (0.355 \pm 0.003)$ and $\beta(q_{x_2} = \infty) = (0.433 \pm 0.005)$.}
    \label{fig:pos_coher}
\end{figure}

Looking at the first direction coherence when using the entangling DTQW, $q_{x_2} = 0.5$, we see that our results Fig.~\ref{fig:pos_coher} reproduces a previously observed behavior, with it showing an initial increase but then decaying and stabilizing around a value $C_{l_1}(\rho_{x_1}(t)) \approx 0.15$. As one increases the amount of randomness in the step sizes of the second direction, the decay rate of the first direction position coherence increases despite the greater initial increase, eventually reaching a stable minimum one with $q_{x_2} = \infty$ and coherence around $C_{l_1}(\rho_{x_1}(t)) \approx 0.05$. As expected, this value is very small when compared to the entangling DTQW, which explains the observed behavior of having a Gaussian-like distribution and yet not a completely separable two-dimensional distribution, since the coherence is not zero. Overall, the first direction walker's position coherence time evolution as a function of $q_{x_2}$ is in agreement with the observed behavior of the marginal distributions Fig.~\ref{fig:pos_panel} with $0.6 < q_{x_2} < 2$ providing distributions between the standard DTQW distribution and a Gaussian one.

But why does the walker in the first direction go under a decoherence process? To answer this question we recall the explanation of why the gEQW using a separable coin operator leads 
to a separable two-dimensional distribution in \ref{subsec:e2dgeqw}. As we stated, the unitary 
operator with a separable coin operator can be rewritten as a independent application of a walk 
in the first direction and in the second direction 
$U_t = S_{x_1}^{(t)}C_{x_1}S_{x_2}^{(t)}C_{x_2}$. A walker density matrix in the 2-D model 
evolves through

\begin{equation}
    \rho_{\hat{x}_1,\hat{x}_2}(t) = \mathcal{U}(t,0)\rho_{\hat{x}_1,\hat{x}_2}(0)\mathcal{U}(t,0)^{\dagger}\mbox{ ,}
\end{equation}
where $\mathcal{U}(t,0) = \prod_{j = 0}^t U_{t-j}$.
By performing a partial trace over the second direction degrees we can find the evolution equation 
for the first direction, therefore

\begin{align*}
    \rho_{\hat{x}_1}(t) &= \sum_{x_2,\sigma_2} \bra{x_2,\sigma_2}\mathcal{U}(t,0)\rho_{\hat{x}_1,\hat{x}_2}(0)\mathcal{U}(t,0)^{\dagger}\ket{x_2,\sigma_2}\mbox{ .}
\end{align*}
Given that the walker directions degrees of freedom, position and coin, are initially separable between 
both directions, i.e. $\rho_{\hat{x}_1,\hat{x}_2}(0) = \rho_{\hat{x}_1}(0) \otimes \rho_{\hat{x}_2}(0)$, 
and noting that both $S_{x_1}^{(t)}$ and $C_{x_1}$ commutes with $S_{x_2}^{(t')}$ and $C_{x_2}$ so that
the evolution operator can be rewritten as a product of two operators that acts only in its respective
direction subspace
$\mathcal{U}(t,0) = \mathcal{U}_{\hat{x}_1}(t,0)\mathcal{U}_{\hat{x}_2}(t,0)$

\begin{equation}
    \rho_{\hat{x}_1}(t) = \mathcal{U}_{\hat{x}_1}(t,0)\rho_{\hat{x}_1}(0)\mathcal{U}_{\hat{x}_1}(t,0)^{\dagger}\mbox{ ,}
\end{equation}
where we used the fact that the evolution for the second direction is unitary and therefore preserves the trace.

Now, if we use an entangling coin operator it is not possible to break the unitary operator into independent 
walks in both directions given that does not exists operators $A_{c_1}$ and $A_{c_2}$ such that 
$C_{entang} = A_{c_1} \otimes A_{c_2}$. Because of that, the walker's directions density matrix 
does not stay separable in the hole evolution and the walker's first direction degrees evolves through 
a non-unitary quantum channel

\begin{align*}
    &\rho_{\hat{x}_1}(t) = \sum_{j} E_{j}^t\rho_{\hat{x}_1}(0)(E_{j}^t)^{\dagger}\mbox{ , where}\\
    & E_{j = \{x_2,\sigma_2\}} = \bra{x_2,\sigma_2}\mathcal{U}(t,0)\ket{\psi_{x_2}(0)}\mbox{ ,}
\end{align*}
with $\rho_{\hat{x}_2}(0) = \ketbra{\psi_{\hat{x}_2}}{\psi_{\hat{x}_2}}$. This quantum channel acts as a 
decoherence channel for the first direction walker evolution. Therefore, we can use the gEQW with an 
entangling coin operator as a controllable decoherence channel in which by controlling the 
parameter $q_{x_2}$ we control the degree of decoherence that the other quantum walk goes under 
(for more details  please check the supplementary material Sec. \ref{ap:entang}).

\section{Concluding remarks}
\label{sec:remarks}

In this work we have presented the study of a tunable 2-D disordered quantum walk, the 2D-gEQW .  By making use of its greater number of degrees of freedom when compared to its 1D predecessor, we have considered another class of coin operator, the entangling coin operator, which creates correlations between the coin subsystems through evolution. We have shown that the 2D-gEQW is not just a simple extension of the corresponding 1D model, but is  associated with new phenomena.

Explicitly, we have found that when using different amount of randomness in the steps sizes in both directions, namely $q$ in the $q$-exponential distribution, it is possible to control the degree of spreading of 
one of the directions. Specifically, if we have the standard discrete time quantum walk in the first direction, by varying $q$ of the second direction steps distributions from $0.5$ to $\infty$ its dynamical exponent goes continuously from the ballistic quantum walk spreading $\alpha = 2$ to the classical spreading $\alpha = 1$, leading the position distribution to a Gaussian-like one, nonetheless not completely separable. In this regard, our protocol adds a new building block for a better understanding of the paths to classical-like  behavior in quantum-walk based systems~\cite{mackay2002quantum,brun2003quantum,schreiber2011decoherence,montero2016classical}.
    
Looking at the time evolution of the position coherence in one direction, we have demonstrated that when we increased the amount of randomness in the steps of the other perpendicular direction the coherence decay rate increases as well, reaching a minimum value of $C_{l_1} \approx 0.05$ for $q_{x_2} = \infty$. That explains the observed properties 
of the first direction position distribution and spreading behavior. Consequently, we conclude that the generalized elephant quantum walk together with an entangling coin operator in a two-dimensional setting can serve as a controllable decoherence channel, with $q$ being the tunable parameter. This is an interesting effect that highlights the role of dimensionality in quantum systems.
    

To understand why the two-dimensional distribution in the entangling generalized elephant quantum walk is not separable, we have analyzed the coin, total and subsystems, entanglement entropy. 
We have found that the coin subsystems become completely entangled when using a $q_{x_2} \ne 1/2$, indicating that the generalized elephant quantum walk in the second direction can also effectively transfer correlations from the subcoins to the subcoins and positions. To eliminate the possibility that the coins be entangled with each other we used the negativity as a measure of entanglement in bipartite systems that returned zero in all occasions, showing that the coins are not entangled with each other. In addition to that, the total coin also showed to be completely correlated with the positions in the same cases, a remarkable feature in 2-D settings. 
 
Let us mention a related 2-D model~\cite{oliveira2006decoherence} where the QW spreads in a lattice subject to broken links. They show that by controlling the probability of breaking links in one direction it is possible to engineer, in a certain range, the degree of decoherence in the perpendicular direction. However, in their protocol the QW always ends up spreading in a diffuse way. In our case, the Fig.~\ref{fig:alphax1_x_qx2} shows that our setup is able to produce diffusion as well as superdiffusion.

While there is a substantial number of works analyzing entanglement features of 1D QWs~\cite{salimi2012asymptotic,rohde2013quantum,vieira2013dynamically,orthey2019weak,singh2019accelerated,montero2016classical,buarque2019aperiodic,pires2020genuine,gratsea2020generation,gratsea2020universal,walczak2021parrondo,pires2021negative,zhang2022maximal}, there are far fewer studies on entanglement in 2-D QWs~\cite{annabestani2010asymptotic,yalccinkaya2015two,zeng2017discrete,angles2022quantum}. Thus, this work also makes a contribution in such subject.
In a previous work\cite{naves2022} the entanglement generation in the 1D 
generalized elephant quantum walk was extensively studied, where it was demonstrated that this type of walk can generate maximally entangled coin states for almost all initial parameters. Adding the above mentioned results, the present work shows that the generalized elephant quantum walk also has the potential to generate highly entangled coin states in 2-D settings, while also maintaining the ability of controlling the walker spreading rate, something that in the context of other types of random quantum walks is not possible. Therefore, we can affirm that the programmable nature of the generalized elephant quantum walk is enriched when taking into account the results presented here, being also useful in the context of decoherence quantum channels design.
    
In a quantum simulation~\cite{georgescu2014quantum}, a central objective is to design quantum systems that could be programmed to model chosen features of other quantum systems. 
In such research field, quantum walks have been shown to be a versatile platform to simulate a myriad of quantum phenomena~\cite{kadian2021quantum,wang2013physical}. In this work, we have shown a new capability with 2-D quantum walks: without weakening the  entanglement, it is possible to control the  wavepacket spreading in one direction by programming a parameter setup for the spreading in the perpendicular direction.

\begin{acknowledgments}
CBN acknowledges the support from IFSC-USP and from Coordena\c{c}\~{a}o de Aperfei\c{c}oamento de Pessoal de N\'{i}vel Superior - Brasil (CAPES) - Finance Code 001. MAP acknowledges the support by the funding agency FUNCAP. DOSP acknowledges the support by the Brazilian funding agencies CNPq (Grant No. 307028/2019-4), FAPESP (Grant No. 2017/03727-0), and the Brazilian National Institute of Science and Technology of Quantum Information (INCT/IQ). SMDQ acknowledges the financial support from CNPq Grant No. 307028/2019-4).
\end{acknowledgments}

\newpage

\onecolumngrid 

\appendix
\section{Appendix}

\subsection{Separable gEQW evolution features}
\label{ap:sep}

To see that the $2$-D generalized elephant quantum walk using a separable 
coin operator leads to a separable probability distribution, we first need to 
take a closer look in the unitary evolution operator

\begin{align}
    U_t &= S_t(\mathbb{I}_p \otimes C_S)\nonumber \\ &= \sum_{x_1,x_2}\bigg( 
    \ketbra{x_1 + \Delta_t^{(1)}, x_2 + \Delta_t^{(2)}}{x_1,x_2} \otimes 
    \ketbra{\uparrow,\uparrow}{\uparrow,\uparrow} + 
    \ketbra{x_1 + \Delta_t^{(1)}, x_2 - \Delta_t^{(2)}}{x_1,x_2} \otimes 
    \ketbra{\uparrow,\downarrow}{\uparrow,\downarrow} \nonumber \\ &+  
    \ketbra{x_1 - \Delta_t^{(1)}, x_2 + \Delta_t^{(2)}}{x_1,x_2} \otimes 
    \ketbra{\downarrow,\uparrow}{\downarrow,\uparrow} + 
    \ketbra{x_1 - \Delta_t^{(1)}, x_2 - \Delta_t^{(2)}}{x_1,x_2} \otimes 
    \ketbra{\downarrow,\downarrow}{\downarrow,\downarrow}\bigg)
    \left(\mathbb{I}_{x_1} \otimes \mathbb{I}_{x_2} \otimes C_2^{(1)} \otimes 
    C_2^{(2)} \right)\mbox{ .}
\end{align}
The shift operator part can be rewritten as a product of two 
operations

\begin{align}
    S_t &= \sum_{x_1,x_2} \bigg( \ketbra{x_1 + \Delta_t^{(1)}}{x_1} \otimes 
    \mathbb{I}_{x_2} \otimes \ketbra{\uparrow}{\uparrow} \otimes \mathbb{I}_{c_2}
    + \ketbra{x_1 - \Delta_t^{(1)}}{x_1} \otimes 
    \mathbb{I}_{x_2} \otimes \ketbra{\downarrow}{\downarrow} \otimes \mathbb{I}_{c_2}\bigg)\cdot \nonumber \\ &\bigg( \mathbb{I}_{x_1} \otimes \ketbra{x_2 + \Delta_t^{(2)}}{x_2}
     \otimes \mathbb{I}_{c_1} \otimes \ketbra{\uparrow}{\uparrow} + 
     \mathbb{I}_{x_1} \otimes \ketbra{x_2 - \Delta_t^{(2)}}{x_2}
     \otimes \mathbb{I}_{c_1} \otimes \ketbra{\downarrow}{\downarrow} \bigg)\mbox{ ,}
\end{align}
where $\mathbb{I}_{c_1}$, $\mathbb{I}_{c_2}$ are the identity operators in the 
first and second direction subcoin spaces, respectively. By also rewriting the 
coin operator as a product of two coin operations, each in its respective coin 
subspace, the unitary operator can be rewritten as

\begin{align}
    &U_t  = S_{x_1}^{(t)}S_{x_2}^{(t)}C_{x_1}C_{x_2}\mbox{ ,}
\end{align}
where
\begin{align}
    &S_{x_1}^{(t)} = \sum_{x_1}\bigg( \ketbra{x_1 + \Delta_t^{(1)}}{x_1} \otimes 
    \mathbb{I}_{x_2} \otimes \ketbra{\uparrow}{\uparrow} \otimes \mathbb{I}_{c_2}
    + \ketbra{x_1 - \Delta_t^{(1)}}{x_1} \otimes 
    \mathbb{I}_{x_2} \otimes \ketbra{\downarrow}{\downarrow} \otimes \mathbb{I}_{c_2}\bigg) \\
    \nonumber \\
    &S_{x_2}^{(t)} = \sum_{x_2}\bigg( \mathbb{I}_{x_1} \otimes \ketbra{x_2 + \Delta_t^{(2)}}{x_2}
     \otimes \mathbb{I}_{c_1} \otimes \ketbra{\uparrow}{\uparrow} + 
     \mathbb{I}_{x_1} \otimes \ketbra{x_2 - \Delta_t^{(2)}}{x_2}
     \otimes \mathbb{I}_{c_1} \otimes \ketbra{\downarrow}{\downarrow} \bigg) \\
    \nonumber \\
    &C_{x_1} = \left(\mathbb{I}_{x_1} \otimes \mathbb{I}_{x_2} \otimes C_{2}^{(1)}
             \otimes \mathbb{I}_{c_2}\right) \\
    \nonumber \\
    &C_{x_2} = \left(\mathbb{I}_{x_1} \otimes \mathbb{I}_{x_2} \otimes 
             \mathbb{I}_{c_1} \otimes C_{2}^{(2)}\right)\mbox{ .}
\end{align}

Every operator that acts only on the first direction subspaces, position 
and subcoin, commutes with the operators which acts only in the other direction,
i.e. 
\begin{align}
    &[S_{x_1}^{(t)},S_{x_2}^{(t')}] = [S_{x_1}^{(t)},C_{x_2}] = 0 
    \label{eq:ap_com1}\\
     &[C_{x_1},S_{x_2}^{(t')}] = [C_{x_1}, C_{x_2}] = 0\mbox{ , }\forall t,t'\mbox{ .}
    \label{eq:ap_com2}
\end{align}
Therefore we can rewrite the one-step unitary operator as

\begin{equation}
    U_t = S_{x_1}^{(t)}C_{x_1}S_{x_2}^{(t)}C_{x_2}\mbox{ ,}
    \label{eq:ap_sepev}
\end{equation}
showing that at every time a $2$-D separable gEQW step can be rewritten as two 
independent steps in each direction.

Now, to show that the distribution is separable, let 
$\rho_{\hat{x}_1,\hat{x}_2}(0) = 
\rho_{\hat{x}_1}(0) \otimes \rho_{\hat{x}_2}(0)$ be the total walker state in 
the initial time step. The total state at a future time $t$ will be given by 

\begin{equation}
    \rho_{\hat{x}_1,\hat{x}_2}(t) = \mathcal{U}(t,0)
    \rho_{\hat{x}_1,\hat{x}_2}(0)\mathcal{U}(t,0)^{\dagger}\mbox{ ,}
    \label{eq:ap_rhoev}
\end{equation}
with $\mathcal{U}(t,0) = \prod_{j = 0}^t U_{t-j}$. The position probability distribution is 
obtained through the partial trace over the coin degrees of freedom and by 
applying a projective measurement $P_{x_1,x_2} = \ketbra{x_1,x_2}{x_1,x_2}$

\begin{equation}
    \mbox{Pr}(x_1,x_2)(t) = \mbox{tr}(P_{x_1,x_2}\rho_{x_1,x_2}(t)P_{x_1,x_2}^{\dagger})
    \mbox{ ,}
\end{equation}
with $\rho_{x_1,x_2}(t) = \mbox{tr}_{c_1,c_2}(\rho_{\hat{x}_1,\hat{x}_2}(t))$. 
Given that the evolution operator applies independent steps to each direction, 
consequence of the separable coin operator, the coin's subsystems will remain 
separable through the entire evolution and the remaining position density 
operator also. Note that $S_{x_1}^{(t)}C_{x_1}$ and $S_{x_2}^{(t)}C_{x_2}$ acts 
only in $\hat{x}_1$ and $\hat{x}_2$, respectively. Also, remember that these 
operators commute between each other, even in different time steps. 
Consequently, the total evolution does not create correlations between the 
directions degrees of freedom and the total probability distribution will be

\begin{align}
    \mbox{Pr}(x_1,x_2)(t) &= \mbox{tr}(P_{x_1,x_2}\rho_{x_1,x_2}(t)P_{x_1,x_2}^{\dagger})
    = \mbox{tr}(P_{x_1,x_2}(\rho_{x_1}(t)\otimes\rho_{x_2}(t))P_{x_1,x_2}^{\dagger}) 
    \nonumber \\
    &= \mbox{tr}(\ketbra{x_1}\rho_{x_1}(t)\ketbra{x_1} \otimes 
     \ketbra{x_2}\rho_{x_2}(t)\ketbra{x_2})
     = \mbox{Pr}(x_1,t)\mbox{Pr}(x_2,t)\mbox{ ,}
    \label{eq:ap_sepprob}
\end{align}
proving that the separable gEQW position distribution is separable.

For completeness, let us address the task of finding the reduced dynamics of a 
given walker direction state. For that, first, we can rewrite the unitary 
evolution $\mathcal{U}(t,0)$ in a more appropriate form

\begin{align}
    \mathcal{U}(t,0) &= \prod_{j = 0}^t U_{t-j} = U_t U_{t-1} \dots U_0 = 
    (S_{x_1}^{(t)}C_{x_1}S_{x_2}^{(t)}C_{x_2})
    (S_{x_1}^{(t-1)}C_{x_1}S_{x_2}^{(t-1)}C_{x_2}) \dots 
    (S_{x_1}^{(0)}C_{x_1}S_{x_2}^{(0)}C_{x_2}) \nonumber \\
    &=(S_{x_1}^{(t)}C_{x_1})(S_{x_2}^{(t)}C_{x_2})
    (S_{x_1}^{(t-1)}C_{x_1})(S_{x_2}^{(t-1)}C_{x_2}) \dots 
    (S_{x_1}^{(0)}C_{x_1})(S_{x_2}^{(0)}C_{x_2})\mbox{ .}
\end{align}
Remembering the commutation relations Eqs.(\ref{eq:ap_com1}) and (\ref{eq:ap_com2}), 
we see that it is possible to permute the operators that acts only in one 
direction with the ones that acts only in the other so that

\begin{align}
    \mathcal{U}(t,0) &= 
    [(S_{x_1}^{(t)}C_{x_1})(S_{x_1}^{(t-1)}C_{x_1})\dots(S_{x_1}^{(0)}C_{x_1})]
    [(S_{x_2}^{(t)}C_{x_2})(S_{x_2}^{(t-1)}C_{x_2}) \dots(S_{x_2}^{(0)}C_{x_2})] 
    \nonumber \\
    &= \mathcal{U}_{\hat{x}_1}(t,0)\mathcal{U}_{\hat{x}_2}(t,0)\mbox{ ,}
    \label{eq:ap_totsepev}
\end{align}
showing that the whole evolution can be described by an independent evolution in 
the second direction and then an independent evolution in the first.

To find the quantum walker's first direction state evolution we need to realize 
a partial trace over the second directions degrees of freedom 
$\mathcal{H}_{x_2}$ and $\mathcal{H}_{c_2}$

\begin{equation}
\rho_{\hat{x}_1}(t) = \mbox{tr}_{\hat{x}_2}\left(\mathcal{U}(t,0)
    \rho_{\hat{x}_1,\hat{x}_2}(0)\mathcal{U}(t,0)^{\dagger}\right)\mbox{ ,}
\end{equation}
now using Eq. (\ref{eq:ap_totsepev})

\begin{align}
    \rho_{\hat{x}_1}(t) &= \mbox{tr}_{\hat{x}_2}\left(\mathcal{U}_{\hat{x}_1}(t,0)
    \mathcal{U}_{\hat{x}_2}(t,0)\rho_{\hat{x}_1,\hat{x}_2}(0)
    \mathcal{U}_{\hat{x}_2}(t,0)^{\dagger}
    \mathcal{U}_{\hat{x}_1}(t,0)^{\dagger}\right) \nonumber \\
    &= \sum_{x_2,\sigma_2} \bra{x_2,\sigma_2}\left(\mathcal{U}_{\hat{x}_1}(t,0)
    \mathcal{U}_{\hat{x}_2}(t,0)\rho_{\hat{x}_1,\hat{x}_2}(0)
    \mathcal{U}_{\hat{x}_2}(t,0)^{\dagger}
    \mathcal{U}_{\hat{x}_1}(t,0)^{\dagger}\right)\ket{x_2,\sigma_2}\mbox{ ,}
    \label{eq:ap_r1ev}
\end{align}
where $\sigma_2 = \{\uparrow,\downarrow\}$. Let again that the initial state be 
separable between the directions. Knowing that the operator $\bra{x_2,\sigma_2}$ 
only is going to act on the identities in $\mathcal{H}_{x_2}$ and 
$\mathcal{H}_{c_2}$ of $\mathcal{U}_{\hat{x}_1}(t,0)$, we can see its effect 
over the evolution operator noting that, for instance

\begin{equation}
    \sum_{x_2}\bra{x_2}\mathbb{I}_{x_2} = \sum_{x_2,x_2'}\bra{x_2}\ketbra{x_2'} 
    = \delta_{x_2,x_2'}\bra{x_2'} = \sum_{x_2}\bra{x_2}\mbox{ ,}
\end{equation}
such that the action of the partial trace over $\mathcal{H}_{x_2}$ passes 
through $\mathcal{U}_{\hat{x}_2}(t,0)$ and the same goes for the partial trace over 
$\mathcal{H}_{c_2}$. Consequently

\begin{align}
    \rho_{\hat{x}_1}(t) &= \sum_{x_2,\sigma_2}\left(\mathcal{U}_{\hat{x}_1}(t,0)
    \bra{x_2,\sigma_2}\mathcal{U}_{\hat{x}_2}(t,0)
    \rho_{\hat{x}_1}(0) \otimes\rho_{\hat{x}_2}(0)
    \mathcal{U}_{\hat{x}_2}(t,0)^{\dagger}\ket{x_2,\sigma_2}
    \mathcal{U}_{\hat{x}_1}(t,0)^{\dagger}\right) = \nonumber \\
    &= \mathcal{U}_{\hat{x}_1}(t,0)\rho_{\hat{x}_1}(0)\mathcal{U}_{\hat{x}_1}(t,0)^{\dagger}\sum_{x_2,\sigma_2}\left(
    \bra{x_2,\sigma_2}\mathcal{U}_{\hat{x}_2}(t,0)\rho_{\hat{x}_2}(0)
    \mathcal{U}_{\hat{x}_2}(t,0)^{\dagger}\ket{x_2,\sigma_2}
    \right) \nonumber \\
    &= \mathcal{U}_{\hat{x}_1}(t,0)\rho_{\hat{x}_1}(0)\mathcal{U}_{\hat{x}_1}(t,0)^{\dagger}\mbox{tr}(\rho_{\hat{x}_2}(t)) = \mathcal{U}_{\hat{x}_1}(t,0)\rho_{\hat{x}_1}(0)\mathcal{U}_{\hat{x}_1}(t,0)^{\dagger}\mbox{ ,}
\end{align}
where in the third line we used the fact that the trace of 
the second direction density matrix is preserved since 
$\mathcal{U}_{\hat{x}_2}(t,0)$ is unitary. This show us that in the 
separable gEQW the evolution of the first direction can be obtained through the 
elimination of the second direction's degrees and that it is simply given by 
$\mathcal{U}_{\hat{x}_1}(t,0)$.

\subsection{Entangling coin operator gEQW evolution features}
\label{ap:entang}

When using a non-separable coin operator, such as Eq. (\ref{eq:2d_entang_coin}), 
the unitary operator cannot be broken into the application of unitary and 
independent quantum walk steps as in Eq. (\ref{eq:ap_sepev}), since the coin 
operator now acts jointly on both coin subspaces. This means that even if we 
start with a separable walker state in both directions, the unitary evolution 
does not necessarily will maintain this property, such that 
$\rho_{\hat{x}_1,\hat{x}_2}(t) \ne \rho_{\hat{x}_1}(t) \otimes \rho_{\hat{x}_2}(t)$. 
Therefore Eq. (\ref{eq:ap_sepprob}) will not be obeyed

\begin{equation}
    \mbox{Pr}(x_1,x_2)(t) \ne \mbox{Pr}(x_1,t)\cdot\mbox{Pr}(x_2,t)\mbox{ .}
\end{equation}

The entangling coin operator Eq. (\ref{eq:2d_entang_coin}) creates correlations 
between the coins through the use of a controlled NOT gate, that uses the first 
degree of freedom as a control and the second as a target. Its action on the 
basis states is $\mbox{CNOT}\{\ket{\uparrow,\uparrow}, \ket{\uparrow,\downarrow}, 
              \ket{\downarrow,\uparrow},\ket{\downarrow,\downarrow}\} = 
             \{\ket{\uparrow,\uparrow}, \ket{\uparrow,\downarrow}, 
              \ket{\downarrow,\downarrow},\ket{\downarrow,\uparrow}\}$.
Now if we have the following superposition state 
$(\ket{\uparrow} + \ket{\downarrow})\ket{\downarrow}$, through the action of the 
CNOT gate the resulting state is an entangled one $(\ket{\uparrow,\downarrow} + \ket{\downarrow,\uparrow})$. 
After the application of $C_{entang}$, the correlations are then transferred to the position-coin
subsystems through the shift operation while also correlating the position degrees, as we saw in 
Figs.~\ref{fig:dtqw_td_entang} and \ref{fig:eqw_td_entang}. 

Because of this inseparability of the unitary operator, the evolution for the 
quantum walker in a given direction will not be unitary anymore. Consider again 
Eq. (\ref{eq:ap_r1ev}) that applying the partial trace over the second direction 
degrees of freedom provides the walker state in the first direction. 
Supposing again that $\rho_{\hat{x}_1,\hat{x}_2}(0) = \rho_{\hat{x}_1}(0) \otimes \rho_{\hat{x}_2}(0)$ 
and that $\rho_{\hat{x}_2}(0) = \ketbra{\psi_{\hat{x}_2}(0)}$, then, given that 
the unitary operator is not separable anymore, we find 
a set of Kraus operators that defines the walker's first direction evolution 

\begin{align}
    \rho_{\hat{x}_1}(t) &=  \sum_{x_2,\sigma_2} \bra{x_2,\sigma_2}\mathcal{U}(t,0)(
    \rho_{\hat{x}_1}(0)\otimes \ketbra{\psi_{\hat{x}_2}(0)})\mathcal{U}(t,0)^{\dagger}\ket{x_2,\sigma_2} \nonumber \\
    &= \sum_{j}E_j(t,0)\rho_{\hat{x}_1}(0)E_j^{\dagger}(t,0)\mbox{ ,}
\end{align}
with $j = \{x_2,\sigma_2\}$ and each Kraus operator given by

\begin{equation}
    E_{j = x_2,\sigma_2}(t,0) = \bra{x_2,\sigma_2}\mathcal{U}(t,0)\ket{\psi_{\hat{x}_2}(0)}\mbox{ .}
\end{equation}
For the evolution be unitary only one Kraus operator should appear, something 
that only is possible if the total unitary operator can be separated into two 
operations each in its respective subspace, as we saw in the case of a separable 
coin operator Sec. \ref{ap:sep}.

\newpage

\twocolumngrid 

\FloatBarrier
\providecommand{\noopsort}[1]{}\providecommand{\singleletter}[1]{#1}

\end{document}